\documentstyle{l-aa}
\input epsf.tex
\newcommand {\Mpc}   {\mbox{$h^{-1}$ Mpc \,}}

\newcommand{\mincir}{\raise -2.truept\hbox{\rlap{\hbox{$\sim$}}\raise5.truept
\hbox{$?$}\ }}
\newcommand{\gr}{\kern 2pt\hbox{}^\circ{\kern -2pt K}} 
\newcommand{\magcir}{\raise -2.truept\hbox{\rlap{\hbox{$\sim$}}\raise5.truept
\hbox{$?$}\ }}

\newcommand{\Om}{\Omega}

\newcommand{\be}{\begin{equation}}
\newcommand{\ee}{\end{equation}}
\newcommand{\bea}{\begin{eqnarray}}
\newcommand{\eea}{\end{eqnarray}}

\newcommand{\si}{\sigma}
\newcommand{\La}{\Lambda}
\newcommand{\etal}{{et al.}}
\begin{document}

\thesaurus{
       (12.03.4;  
        12.04.1;  
        12.12.1;  
Universe
        11.03.1;  
          }

\title{Cosmological parameters from large scale structure observations}

\author{B.~Novosyadlyj \inst{1}, R.~Durrer \inst{2},
S.~Gottl\"ober \inst{3}, V.N.~Lukash \inst{4}, S.~Apunevych \inst{1}}

\offprints{Bohdan Novosyadlyj}

\institute{Astronomical Observatory of L'viv State University, Kyryla and
Mephodia str.8, 290005, L'viv, Ukraine
\and Department de Physique Th\'eorique, Universit\'e de Gen\`eve,
Quai Ernest Ansermet 24, CH-1211 Gen\`eve 4, Switzerland
\and Astrophysikalisches Institut Potsdam, An der Sternwarte 16,
D-14482 Potsdam, Germany
\and Astro Space Center of Lebedev Physical Institute of RAS,
Profsoyuznaya 84/32, 117810 Moscow, Russia}

\date{Received \dots; accepted \dots}
\maketitle
\markboth{Cosmological parameters from LSS}{ Novosyadlyj et al.}

\begin{abstract}
The possibility of determining cosmological parameters on the basis of a wide set
of observational data including the Abell-ACO cluster power spectrum and mass
function, peculiar velocities of galaxies, the distribution of Ly-$\alpha$ clouds
and CMB temperature fluctuations is analyzed.  Using a $\chi^2$ minimization method,
assuming $\Omega_{\Lambda}+\Omega_{\rm{matter}} =1 $ and no contribution from gravity
waves, we show that this data set determines quite precisely the values of the
spectral index $n$ of the primordial power spectrum, baryon, cold dark matter 
and massive neutrino density $\Omega_b$, $\Omega_{cdm}$ and $\Omega_{\nu}$ respectively, 
the Hubble constant $h\equiv H_0/(100$km/s/Mpc) and the value of the cosmological constant,
$\Omega_{\Lambda}$ .

Varying all parameters, we found that a tilted $\Lambda$MDM model with one sort
of massive neutrinos and the parameters $n=1.12\pm 0.10$, $\Omega_m=0.41\pm 0.11$
($\Omega_{\Lambda}=0.59\pm0.11$), $\Omega_{cdm}=0.31\pm 0.15$,
$\Omega_{\nu}=0.059\pm 0.028$, $\Omega_b=0.039\pm 0.014$ and $h=0.70\pm 0.12$
matches  observational data best.

$\Omega_{\nu}$ is higher for more species of massive neutrinos, $\sim 0.1$
for two  and $\sim 0.13$ for three species. $\Omega_m$ raises by $\sim 0.08$
and $\sim 0.15$ respectively.

The 1$\sigma$ (68.3\%) confidence limits on each cosmological parameter, which are
obtained by marginalizing over the other parameters, are $0.82\le n\le1.39$,
$0.19\le\Omega_m\le 1$ ($0\le\Omega_{\Lambda}\le 0.81$),
$0\le\Omega_{\nu}\le 0.17$, $0.021\le \Omega_b\le 0.13$ and
$0.38\le h\le 0.85$ $1.5\le b_{cl}\le 3.5$. Here $b_{cl}$ is the
cluster bias parameter. The best-fit parameters for 31 models
which are inside of $1\sigma$ range of the best model are presented (Table 4).

Varying only a subset of parameters and fixing the others changes 
the results. In
particular, if a pure matter model ($\Omega_m=1$) is assumed, MDM with
$\Omega_{\nu}=0.22\pm 0.08$, three species of massive neutrinos and low
$h=0.47\pm 0.05$  matches the observational data best. If a low density Universe
$\Omega_m=0.3$ is assumed, a $\Lambda$CDM model without hot dark matter
and high $h=0.71$ matches
the observational data best. If the primordial power spectrum is scale invariant
($n \equiv 1$) a low density Universe ($\Omega_{m}=0.45\pm 0.12,~~h=0.71\pm 0.13$)
with  very little hot dark matter ($\Omega_{\nu}=0.04\pm 0.03$, 
$N_{\nu}=1$) becomes the best fit.

It is shown also that observational data set used here rules out
the class of CDM models with $h\ge 0.5$, scale invariant primordial power
spectrum, zero cosmological constant and spatial curvature at very
high confidence level, $>99.99\%$.
The corresponding class of MDM models are ruled out at $\sim 95\%$ C.L.

\end{abstract}

 \keywords{Large Scale Structure: cosmological models,
power spectrum, cosmological parameters}

\section{Introduction}

Observations of the large scale structure (LSS) of the Universe
carried out during the last years and coming up from current experiments and
observational programs allow to determine  the parameters
of cosmological models and the nature of dark matter more precisely. The usual
cosmological paradigm - a scale free power spectrum of scalar
primordial perturbations which evolve in a multicomponent medium to
form the large scale structure of the Universe - is compatible with
the observed cosmic microwave background (CMB) temperature
fluctuations.  Most
inflationary scenarios predict a scale free primordial power spectra
of scalar density fluctuations $P(k)\sim k^{n}$ with arbitrary $n$ as
well as gravity waves which contribute to the power spectrum of CMB
temperature fluctuations $({\Delta T\over T})_{\ell}$ at low spherical
harmonics. But models with a minimal number of free parameters, such
as the scale invariant ($n=1$) standard cold dark matter model
(SCDM) or the standard mixed (cold plus hot) dark matter model (SMDM)
only marginally match observational data. Better agreement between
predictions and observational data can be achieved in models with a
larger number of parameters: cold dark matter (CDM) or mixed dark
matter (MDM) with baryons, a tilted primordial power spectra, spatial
curvature ($\Omega_k$), a cosmological constant 
($\Omega_{\Lambda}$) and a tensor
contribution to the CMB anisotropy power spectrum.

The neutrino oscillations discovered recently in the Super-Kamiokande
experiment (\cite{fu98}) show that at least one species of weakly
interacting neutrinos have non-zero rest mass. Assuming that the
larger one of them is about  $m_{\nu} \simeq
\sqrt{\delta m_{\nu}^2} \approx 0.07$ eV we find $\Omega_{\nu} \approx 7.4
\times 10^{-4}
N_\nu/h^2$.  It is also possible that one, two or three species have
masses in the eV range and give appreciable contribution to the dark
matter content of the Universe. 

The presence of rich clusters of galaxies at $z\approx 0.54,\;0.55,\;0.8$
(\cite{bah98}) indicates a low matter density. 

In this work we do not include into
our analysis the recent observations of distant supernovae
(\cite{per98,rie98}). The SNeIa measurements support a positive
cosmological constant.  Assuming a flat Universe,
$\Omega_{\Lambda}+\Omega_m=1$, a value of $\Omega_{\Lambda}\sim 0.7$
is preferred (see also the review by \cite{bah99}), but, in agreement
with \cite{vkn98,pr98}, we find that on the basis of LSS data alone, a
non-vanishing cosmological constant is preferred within the class of
models analyzed in this work.

Another approach based on the
search of best-fit cosmological parameters in open and critical
density CDM and $\Lambda$CDM models without gravitational waves 
for the total combination of
observational data on CMB anisotropy has been carried out by
\cite{lin97}. But the CMB data set corresponds to very large scales
($\ge 100h^{-1}$Mpc) and it is not sufficiently sensitive to the existence
of a HDM component.  The power spectra of density
fluctuations obtained from the spatial distribution of Abell-ACO clusters
(\cite{ein97,ret97}), APM, CfA and IRAS galaxy surveys
(\cite{ein99} and references therein) are extended to smaller scales
up to galaxy scales which are below the neutrino free streaming
scale. On small scales constraints are obtained from absorption
features in quasar spectra known as the Ly-$\alpha$ forest
(\cite{gn98,cr98}).

The determination of cosmological parameters from some observations
 of the LSS of the Universe was carried out in many
papers (e.g. \cite{atr97,lin97,Teg99,bri99,nov99} and references therein).
Recently, {\cite{bri99}} have analyzed the cluster abundances,
CMB anisotropies and IRAS observations to optimize the four parameters
($\Omega_m$, $h$, $\sigma_8$, and $b_{\rm IRAS}$ in a open CDM model.
{\cite{atr97}} use the
cluster power spectrum together with data of the Saskatoon experiment
to discuss the possible existence of a built-in scale in the
primordial power spectrum. In this paper a total of 23
measurements from sub-galaxy scales (Ly-$\alpha$ clouds) over cluster
scales up to the horizon scale (CMB quadrupole) are used to
determine seven cosmological parameters.

Clearly, it is possible that the 'correct cosmological model' is not
one of those analyzed in this paper. If the data are good enough this
can in principle be decided by a $\chi^2$-test. As long as we find a
model within the family of models studied here with an acceptable
value of $\chi^2$, we have no compelling reason to consider other models.

 In view of the growing body of observational data, we
want to discuss the quantitative differences between theory
and observations for the entire class of available models by varying
all the input parameters such as the tilt of the primordial spectrum, $n$,
the density of cold dark matter, $\Omega_{cdm}$, hot dark matter,
$\Omega_{\nu}$, and baryons, $\Omega_b$, the vacuum energy or
cosmological constant, $\Omega_{\Lambda}$, and the Hubble parameter
$h$, to find the values which agree best with observations of LSS on
all scales (or even to exclude a whole family of models). Here we
restrict ourselves to the analysis of spatially flat cosmological
models with $\Omega_{\Lambda}+\Omega_m=1$ ($\Omega_k=0$), where
$\Omega_m=\Omega_{cdm}+ \Omega_b+\Omega_{\nu}$, and to an inflationary
scenario without tensor mode. We also neglect the effect of a possible
early reionization which could reduce the amplitude of the first
acoustic peak in the CMB anisotropy spectrum.

The reason for the restriction of flat models is mainly
numerical. However, the new CMB anisotropy data from the Boomerang
experiment actually strongly favors spatially flat universes~(\cite{mel99b}).
Neglecting the tensor mode which affects the normalization and the
height of the first acoustic peak is motivated by the work of
\cite{Teg99}, who found that CMB anisotropy data prefer no or a small
tensor component, however there are also arguments in favor of
the importance of the tensor mode (\cite{arc98,mel99}). Furthermore, 
since the LSS data used in this paper
disfavors very blue spectra, the high acoustic peak indicates that
reionization cannot be substantial for  the class of models analyzed
in this paper. Hence we set the optical depth $\tau=0$.

The outline of this paper is as follows: In Sect.~2 we describe the
observational data which are used. The method of parameter
determination and some tests are described in
Sect.~3. We present the results obtained under different assumptions
about the parameter ranges in Sect. 4. A discussion of our results and
the conclusions are given in Sects. 5 and 6 respectively.

\section{The experimental data set}

\subsection{The Abell-ACO cluster power spectrum}

\begin{figure}[ht]
\epsfxsize=9truecm
\epsfbox{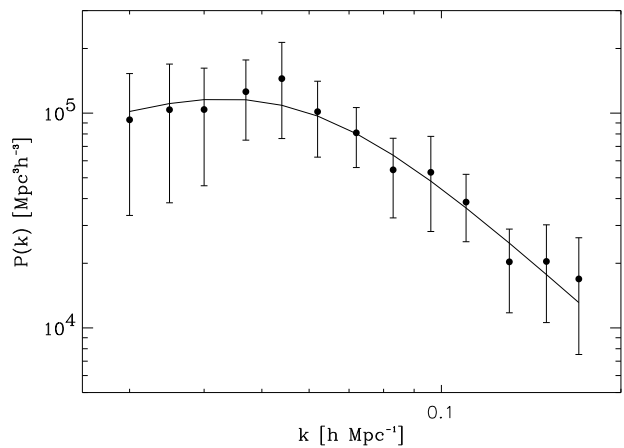}
\caption{The Abell-ACO power spectrum by Retzlaff \etal ~ 1998. The
solid line is the best fit according to Eqs. 1 and 2.}
\label{PSfig}
\end{figure}

One might expect that the most favorable data for the determination of
cosmological parameters are power spectra constructed from the
observed distribution of galaxies. But the power spectra of galaxies
obtained from the two-dimensional  APM survey
(e.g. \cite{mad96,tad96}, and references therein), the CfA redshift survey
(\cite{vog92,par94}), the IRAS survey (\cite{sau92}) and/or from the Las
Campanas Redshift Survey (\cite{dc94,lan96}) differ both in the
amplitude and in the behavior near the maximum.  Moreover, nonlinear
effects on small scale must be taken into account in their analysis.
For these reasons we do not include galaxy power spectra for the
determination of parameters in this work. Here, we use the power
spectrum of Abell-ACO clusters (\cite{ein97,ret97}) as observational
input. This power spectrum is measured in the range
$0.03h/$Mpc$\le k\le 0.2h/$Mpc. The cluster power spectrum is biased with
respect to the dark matter distribution.  We assume that the bias is
linear and scale independent in the range of scales considered. The
position of the maximum ($k_{max}\approx 0.05h/$Mpc) and the slope at
lower and larger scales are sensitive to the baryon content $\Omega_{b}$,
the Hubble constant $h$, the neutrino mass $m_{\nu}$ and the number of
species of massive neutrinos $N_{\nu}$ (\cite{nov99}).
The Abell-ACO cluster power spectrum  $\tilde P_{A+ACO}(k_j)$ (here
and in the following a tilde denotes observed quantities) has been
taken from \cite{ret97}. We present 13 values of $\tilde
P_{A+ACO}(k_j)$ and the $1\sigma$ errors in Table 1 and in Fig.~\ref{PSfig}.
In a first step, we have assumed that the 13 points
in this power spectrum given below are independent measurements. The
value of $\chi^2$ obtained under this assumption is much smaller than
the number of degrees of freedom (see below). We interpret this as a hint that
the 13 points of  $\tilde P_{A+ACO}$ given in Table~I cannot be
considered as independent measurements. We therefore describe the
power spectrum by three parameters $A$, $k_{\rm bend}$ and $\alpha$
to be of the form
\be
\tilde P_{A+ACO}(k) = {Ak\over 1+(k/k_{\rm bend})^\alpha} ~.
\ee
A fit of the parameters to the observed power spectrum gives
$$ A=(3.78\pm 1.71)\times 10^6,~~~ k_{\rm bend} =0.056\pm 0.015,~$$
\be
\alpha=3.49\pm 0.72.
\ee
In Fig.~\ref{PSfig} we show the observed power spectrum together with the fit.
The cosmological model parameters obtained using the full power spectrum
information or the three parameter fit are in good agreement, but the
latter prescription leads to a more reasonable value of $\chi^2$.

This point is quite important since it illustrates that a small
$\chi^2$ need not mean that the error bars of the data are too large
but it can be due to data points depending only on a few parameters
and therefore not being independent. If
a power spectrum, like the one above can be modeled by 3 parameters,
then, by varying three cosmological parameters, like {\em e.g.} the cluster
bias $b_c$~, the HDM contribution $\Om_\nu$ and the Hubble
parameter, $h$, we can in general (if there is no
degeneracy) fit all three parameters $A$, $k_{\rm bend}$ and $\alpha$
and thereby the entire power spectrum. The number of degrees of
freedom in such a fit is $0$ and not $10$ as one would infer form
 the number  points of the power spectrum.

To make best use of the observational information, we nevertheless use
the full 13 points of the power spectrum to fit the data, but we
assign it $n_F=3$ for the number of degrees of freedom.

\begin{table}[th]
\caption{The Abell-ACO power spectrum by Retzlaff \etal ~ 1998}
\def\onerule{\noalign{\medskip\hrule\medskip}}
\medskip
\begin{center}
\begin{tabular}{|ccc|}
\hline
&&\\
  No &  $k_j$ &     $\tilde P_{A+ACO}(k_j) \pm     \Delta \tilde P $\\
[4pt]
\hline
&&\\
  1 &  0.030    &($9.31   \pm 5.97) \cdot 10^4$ \\
  2 &  0.035    &($1.04   \pm 0.66) \cdot 10^5$ \\
  3 &  0.040    &($1.04   \pm 0.58) \cdot 10^5$ \\
  4 &  0.047    &($1.26   \pm 0.51) \cdot 10^5$ \\
  5 &  0.054    &($1.45   \pm 0.69) \cdot 10^5$ \\
  6 &  0.062    &($1.02   \pm 0.39) \cdot 10^5$ \\
  7 &  0.072    &($8.10   \pm 2.52) \cdot 10^4$ \\
  8 &  0.083    &($5.44   \pm 2.19) \cdot 10^4$ \\
  0 &  0.096    &($5.30   \pm 2.49) \cdot 10^4$ \\
 10 &  0.11     &($3.85   \pm 1.33) \cdot 10^4$ \\
 11 &  0.13     &($2.03   \pm 0.86) \cdot 10^4$ \\
 12 &  0.15     &($2.04   \pm 0.98) \cdot 10^4$ \\
 13 &  0.17     &($1.70   \pm 0.94) \cdot 10^4$ \\ [4pt]
\hline
\end{tabular}
\end{center}
\end{table}

\subsection{CMB data}

We normalize the power spectrum using the COBE 4-year data of CMB
temperature fluctuations (\cite{ben96,lid96,bun97}).
We believe that using all
available experimental data on $\Delta T/T$ on angular scales smaller
than the COBE measurement is not an optimal way for searching of
best-fit parameters because some data points in CMB spectrum
contradict each other. Therefore, we use only the position and
amplitude of the first acoustic peak derived from observational
data as integral characteristics of CMB power spectrum, which are
sensitive to some of the model parameters.

\begin{figure}[th]
\epsfxsize=9truecm
\epsfbox{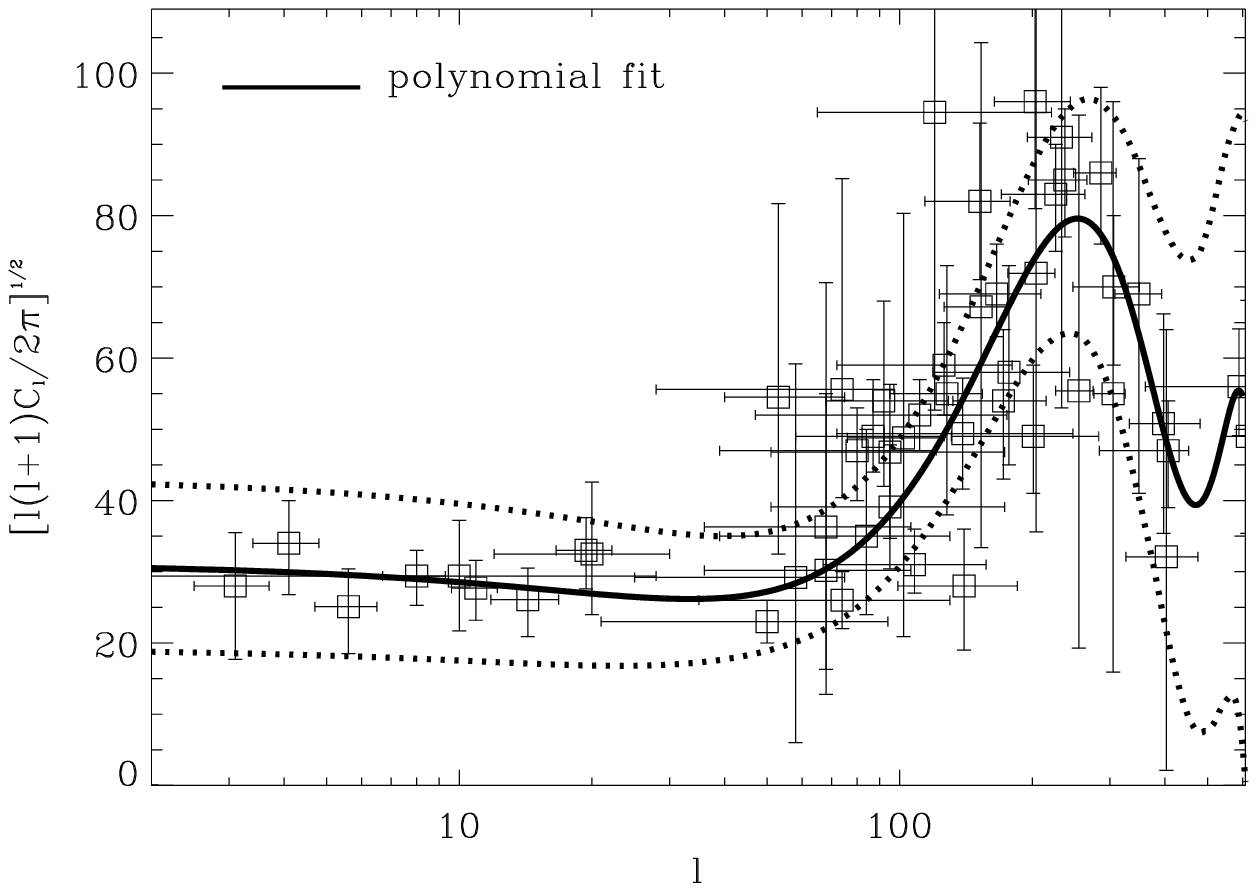}
\caption{Observational data of CMB fluctuation (Table
2) and a sixth order polynomial fit to a power spectrum
(solid line). The dotted lines restrict the space of fitting curves which deviate
from best fit by less than $1\sigma$ ($\Delta\chi^2=47.9$ for 44 degrees of
freedom).}
\label{Clfit}
\end{figure}

To determine the position, $\ell_p$, and amplitude, $A_p$, of the
first acoustic peak we use the set of observational data on CMB
temperature anisotropy given in Table~2 (altogether 51 observational points).  
For each experiment we include the effective harmonic, the
amplitude of the temperature fluctuation at this harmonic, the upper and
lower error in the temperature, and the effective range of the window
in $\ell$-space. In those cases when original papers do not contain 
effective harmonics and band width we have taken them from Max Tegmark's 
CMB data analysis center (\cite{teg}) dated Nov 25 1999.
We fit the experimental data points by a polynomial of 6-th order using the
Levenberg-Marquardt method to determine the position and amplitude of
the first peak: $[l(l+1)C_l/2\pi^2]^{1/2}=\sum_{i=0}^6~a_il^i$. The best-fit
values of the coefficients are: 
$a_0=31.1$, $a_1=-0.309$, $a_2=5.18\times10^{-3}$,
$a_3=9.66\times10^{-6}$, $a_4=-3.68\times10^{-8}$, $a_5=1.22\times10^{-10}$,
$a_6=-9.19\times10^{-14}$ ($\chi^2=62.9$).
The amplitude $A_p$ and position $l_p$ of first acoustic peak
determined from data fitting curve are $79.6\mu \rm K$ and $253$
correspondingly. Our result differs only slightly from the numbers
obtained by \cite{lin97} who found 260 and $88\mu \rm K$.
Fig.~\ref{Clfit} shows the observational data used together with the polynomial
best fit (solid line).

 We estimated the error of  $A_p$ and $l_p$ in the
following way.

\begin{figure}[th]
\epsfxsize=9truecm
\epsfbox{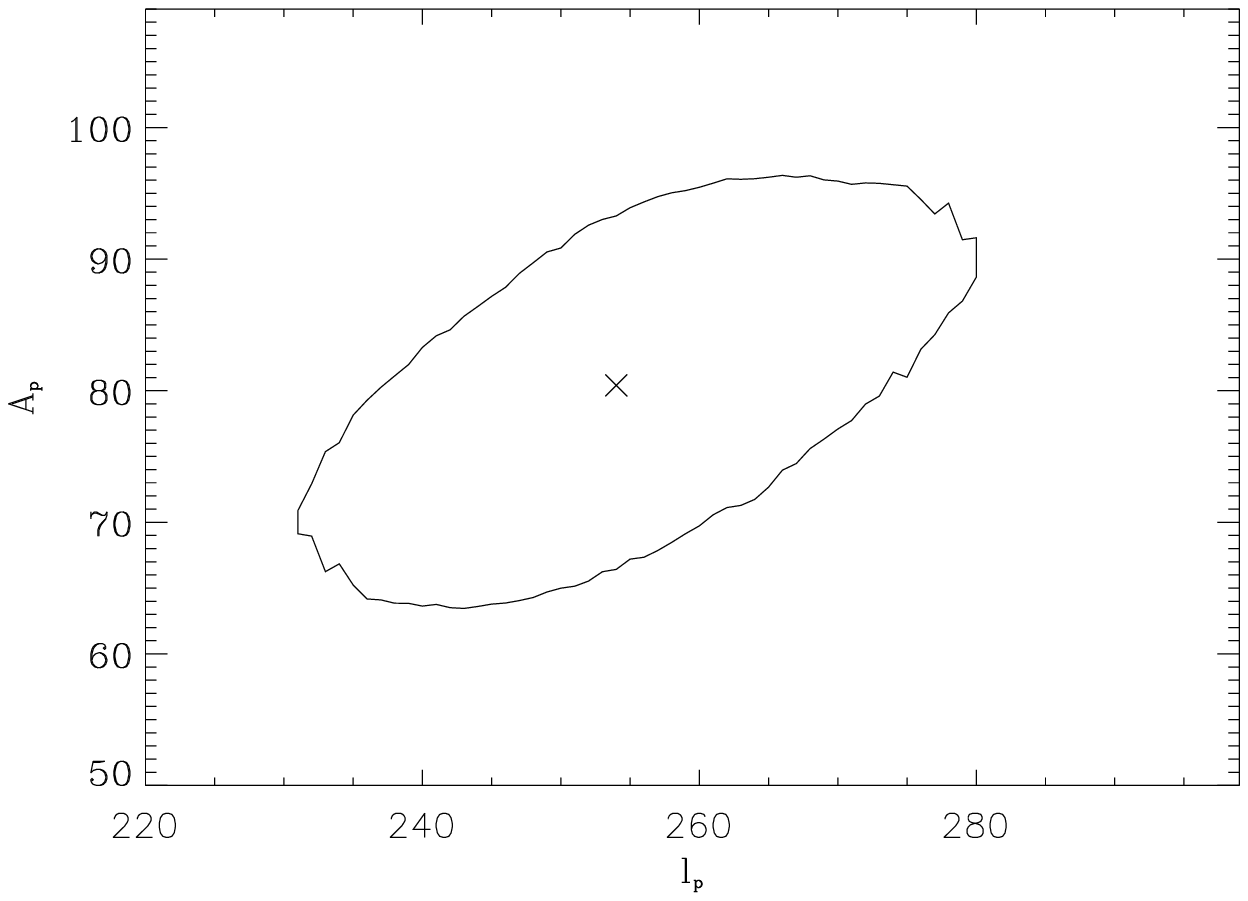}
\caption{The contour of positions $l_p$ and amplitudes $A_p$ of first acoustic peak
which corresponds to the range of fitting curves which are in the 68.3\% range of
probability of point distribution. The box which contains ellipse gives $1\sigma$
errors for $l_p$ and $A_p$. The position $l_p$ and amplitude $A_p$ for best fit
coefficients are shown as a cross (see also in Fig.~\ref{Clfit}).}
\label{dpst}
\end{figure}

 By varying of all coefficients $a_i$  we determine
$\chi^2$-hyper-surface in 7-dimension parameter space which contains
deviations of less than $1\sigma$. If the probability  distribution obeys
Gaussian statistics, this corresponds to a 68.3\% confidence level. It
is well known, that present CMB anisotropy data even on small scales do
not obey Gaussian statistics and thus this procedure is somewhat
arbitrary. However, it can be assumed that this gives us a good
indication for the errors bar in position and amplitude of the first
acoustic peak. For 44 degrees of freedom (51 data points minus 7
parameters)  this
hyper-surface corresponds to $\Delta\chi^2=47.9$. For values of parameters
$a_i$ which have a $\chi^2 < \chi_{min}^2+\Delta\chi^2$
 we calculate the $C_l$'s and find the peak amplitude $A_p$ and the position
$l_p$. They are in the contour line in the  $A_p-l_p$ plane shown in
Fig.~\ref{dpst}. The upper-lower and right-left extremal points indicate
 $1\sigma$ statistical errors: $\Delta A_p^{st}=+17.0,-16.3\mu \rm K$ and
$\Delta l_p^{st}=+28,-22$. Uncertainties of effective harmonics of
each experiment do not influence the error of the amplitude of the
first acoustic peak but must
be take into account for the full error in the peak position,
so that $\Delta l_p=\Delta l_p^{st}+\Delta l_p^{w}$, where last term is
the mean band width around $l_p$. We estimate it as mean width of all
experiments weighted by acoustic peak amplitude
$\Delta l_p^{w}=\sum_{i=1}^{56} (\Delta l)_i\omega_i/\sum_{i=1}^{56}\omega_i$,
where  the weighting factor $\omega_i=[l_i(l_i+1)C_{l_i}/2\pi^2]^{1/2}/A_p$ is
calculated using polynomial fit. This finally leads to
$\Delta l_p^{w}\approx 45.0$.
(Without weighting the  value is $\Delta l_p^{w} \approx 42$). Therefore, the
errors of determination of first acoustic peak amplitude and position are
$\Delta A_p\approx 16.5\mu \rm K$ and $\Delta l_p\approx 70$
respectively. We  use these errors below in our search of
cosmological parameters.

It is interesting to note that no 6th order polynomial fits
the data really well. For our best fit polynomial we obtain
$\chi^2= 62.9$ for $51$ data points and $7$ parameters. The
probability for this polynomial leading to the
observed data is about 1\%. This big $\chi^2$ can have two
origins. First, the probability distribution is non-Gaussian and,
therefore, the probability to obtain this value of  $\chi^2$ is higher
than 1\%. Secondly, some data seems to be
contradictory. For example, if we ignore all the Python V
points we obtain a best fit polynomial with $\chi^2 = 23$ which is
even slightly too low. (Removing of these points does not change essentially
the result amplitude and position of acoustic peak, $\ell_p=256$, $A_p=79.0$
without them). But of course we are not allowed without any
good reason, to leave away some experimental results. It may well be
that Python V is correct and some other experiments are wrong.
Therefore, we adopted this somewhat hand waving way to extract information
from this data. Clearly, a more thorough analysis with true,
non-Gaussian likelihood functions would be in order, which we leave
for the future (see~\cite{bartlett}).

For the comparison of models with the CMB data we use, apart from the COBE
normalization, only the two parameters
obtained by the fitting procedure described above:
the effective harmonic $\ell_p=253\pm70$ of the peak position
and the amplitude of the peak $A_p=79.6\pm16.5\mu \rm K$.
Clearly, this position and height of the
first acoustic peak is not strictly implied by the present data and can
therefore be criticized. In this sense it has to be considered
primarily as a working hypothesis which will be confirmed or
contradicted in the future by more accurate data.
\begin{table*}
\label{ObsCMB}
\center{\caption{Observational data on CMB temperature fluctuations
(in $\mu \rm K$) }}
\begin{center}
\def\onerule{\noalign{\medskip\hrule\medskip}}
\begin{tabular}{|lclllll|}
\hline
 &&&&&&\\
${\ell_{\rm min}}$ & $ {\ell_{\rm eff}}$ &  ${\ell_{\rm max}}$ & $\delta T^{obs}_{\ell_{\rm eff}}$
& ${\it + err}$  &  ${\it - err}$ & Experiment \\ [10pt]
\hline
 &&&&&&\\
${ 2.5 }$&{\bf 3.1} &${ 3.7 }$&{\bf 28.0} & ${+ 7.5}$ & ${-10.3}$ & COBE2, \cite{TegHam}\\
${ 3.4 }$&{\bf 4.1} &${ 4.8 }$&{\bf 34.0} & ${+ 6.0}$ & ${- 7.2}$ & COBE3, \cite{TegHam}\\
${ 4.7 }$&{\bf 5.6} &${ 6.5 }$&{\bf 25.1} & ${+ 5.3}$ & ${- 6.6}$ & COBE4, \cite{TegHam}\\
${ 6.7 }$&{\bf 8  } &${ 9.3 }$&{\bf 29.4} & ${+ 3.6}$ & ${- 4.1}$ & COBE5, \cite{TegHam}\\
${ 2   }$&{\bf 10 } &${ 28  }$&{\bf 29.4} & ${+ 7.8}$ & ${- 7.7}$ & FIRS, \cite{gan94}  $^{*)}$\\
${ 9.6 }$&{\bf 10.9}&${ 12.2}$&{\bf 27.7} & ${+ 3.9}$ & ${- 4.5}$ & COBE6, \cite{TegHam}\\
${ 11.8}$&{\bf 14.3}&${ 16.8}$&{\bf 26.1} & ${+ 4.4}$ & ${- 5.2}$ & COBE7, \cite{TegHam}\\
${ 16.6}$&{\bf 19.4}&${ 22.2}$&{\bf 33.0} & ${+ 4.6}$ & ${- 5.4}$ & COBE8, \cite{TegHam}\\
${ 12  }$&{\bf 20 } &${ 30  }$&{\bf 32.5} & ${+10.1}$ & ${- 8.5}$ & Tenerife, \cite{han97}  $^{*)}$\\
${ 21  }$&{\bf 50 } &${ 94  }$&{\bf 23.0} & ${+ 3.0}$ & ${- 3.0}$ & PythonV1, \cite{coble99}\\
${ 40  }$&{\bf 53 } &${ 75  }$&{\bf 54.5} & ${+27.2}$ & ${-22.0}$ & iac/bartol2, \cite{femen97}\\
${ 36  }$&{\bf 68 } &${ 106 }$&{\bf 30.2} & ${+24.8}$ & ${-17.4}$ & SP91, \cite{gun95}\\
${ 36  }$&{\bf 68 } &${ 106 }$&{\bf 36.3} & ${+34.3}$ & ${-20.0}$ & SP94, \cite{gun95}\\
${ 25  }$&{\bf 58 } &${ 75  }$&{\bf 29.0} & ${+30.0}$ & ${-23.2}$ & Boomerang, \cite{mau99}\\
${ 28  }$&{\bf 74 } &${ 97  }$&{\bf 55.6} & ${+29.6}$ & ${-15.2}$ & BAM, \cite{tuck97}  $^{*)}$\\
${ 35  }$&{\bf 74 } &${ 130 }$&{\bf 26.0} & ${+ 4.0}$ & ${- 4.0}$ & PythonV2, \cite{coble99}\\
${ 39  }$&{\bf 80 } &${ 121 }$&{\bf 47.0} & ${+ 6.0}$ & ${- 7.0}$ & QMap F1+2Ka, \cite{olive98}\\
${ 39  }$&{\bf 84 } &${ 130 }$&{\bf 35.0} & ${+15.0}$ & ${-11.0}$ & MSAM, \cite{msam}\\
${ 58  }$&{\bf 87 } &${ 126 }$&{\bf 49.0} & ${+ 8.0}$ & ${- 5.0}$ & SK1, \cite{net97}\\
${ 68  }$&{\bf 92 } &${ 129 }$&{\bf 54.0} & ${+14.0}$ & ${-12.0}$ & Python1, \cite{pla97}\\
${ 51  }$&{\bf 95 } &${ 173 }$&{\bf 39.1} & ${+ 8.7}$ & ${- 8.7}$ & Argo1, \cite{debern94}  $^{*)}$\\
${ 51  }$&{\bf 95 } &${ 173 }$&{\bf 46.8} & ${+ 9.5}$ & ${-12.1}$ & Argo2, \cite{mas96}  $^{*)}$\\
${ 76  }$&{\bf 102} &${ 125 }$&{\bf 48.8} & ${+31.5}$ & ${-27.9}$ & Boomerang, \cite{mau99}\\
${ 67  }$&{\bf 108} &${ 157 }$&{\bf 31.0} & ${+ 5.0}$ & ${- 4.0}$ & PythonV3, \cite{coble99}\\
${ 47  }$&{\bf 111} &${ 175 }$&{\bf 52.0} & ${+ 5.0}$ & ${- 5.0}$ & QMap F1+2Q, \cite{olive98}\\
${ 65  }$&{\bf 120} &${ 221 }$&{\bf 94.5} & ${+41.8}$ & ${-41.8}$ & IAB, \cite{picc93}  $^{*)}$\\
${ 72  }$&{\bf 126} &${ 180 }$&{\bf 59.0} & ${+ 6.0}$ & ${- 7.0}$ & QMap F1+2Ka, \cite{olive98}\\
${ 95  }$&{\bf 128} &${ 154 }$&{\bf 55.0} & ${+18.0}$ & ${-17.0}$ & TOCO98, \cite{TOCO}\\
${ 72  }$&{\bf 139} &${ 247 }$&{\bf 49.4} & ${+ 7.8}$ & ${- 7.8}$ & MAX, \cite{tan96}  $^{*)}$\\
${ 99  }$&{\bf 140} &${ 185 }$&{\bf 28.0} & ${+ 8.0}$ & ${- 9.0}$ & PythonV4, \cite{coble99}\\
${114  }$&{\bf 152} &${ 178 }$&{\bf 82.0} & ${+11.0}$ & ${-11.0}$ & TOCO98, \cite{TOCO}\\
${126  }$&{\bf 153} &${ 175 }$&{\bf 67.0} & ${+37.1}$ & ${-33.8}$ & Boomerang, \cite{mau99}\\
${123  }$&{\bf 166} &${ 209 }$&{\bf 69.0} & ${+ 7.0}$ & ${- 6.0}$ & SK2, \cite{net97}\\
${119  }$&{\bf 177} &${ 243 }$&{\bf 58.0} & ${+15.0}$ & ${-13.0}$ & Python2, \cite{pla97}\\
${132  }$&{\bf 172} &${ 215 }$&{\bf 54.0} & ${+10.0}$ & ${-11.0}$ & PythonV5, \cite{coble99}\\
${131  }$&{\bf 201} &${ 283 }$&{\bf 49.0} & ${+10.0}$ & ${- 8.0}$ & MSAM, \cite{msam}\\
${164  }$&{\bf 203} &${ 244 }$&{\bf 96.0} & ${+15.0}$ & ${-15.0}$ & PythonV6, \cite{coble99}\\
${176  }$&{\bf 204} &${ 225 }$&{\bf 71.9} & ${+38.7}$ & ${-36.3}$ & Boomerang, \cite{mau99}\\
${170  }$&{\bf 226} &${ 263 }$&{\bf 83.0} & ${+ 7.0}$ & ${- 8.0}$ & TOCO98, \cite{TOCO}\\
${195  }$&{\bf 233} &${ 273 }$&{\bf 91.0} & ${+32.0}$ & ${-38.0}$ & PythonV7, \cite{coble99}\\
${196  }$&{\bf 237} &${ 266 }$&{\bf 85.0} & ${+10.0}$ & ${- 8.0}$ & Sk3, \cite{net97}\\
${226  }$&{\bf 255} &${ 275 }$&{\bf 61.0} & ${+38.7}$ & ${-36.1}$ & Boomerang, \cite{mau99}\\
${248  }$&{\bf 286} &${ 310 }$&{\bf 86.0} & ${+12.0}$ & ${-10.0}$ & SK4, \cite{net97}\\
${276  }$&{\bf 305} &${ 325 }$&{\bf 55.0} & ${+40.9}$ & ${-39.1}$ & Boomerang, \cite{mau99}\\
${247  }$&{\bf 306} &${ 350 }$&{\bf 70.0} & ${+10.0}$ & ${-11.0}$ & TOCO98, \cite{TOCO}\\
${308  }$&{\bf 349} &${ 393 }$&{\bf 69.0} & ${+19.0}$ & ${-28.0}$ & SK5, \cite{net97}\\
${332  }$&{\bf 397} &${ 481 }$&{\bf 50.8} & ${+15.4}$ & ${-15.4}$ & CAT1, \cite{sco96}  $^{*)}$\\
${326  }$&{\bf 403} &${ 475 }$&{\bf 32.0} & ${+31.9}$ & ${-30.0}$ & Boomerang, \cite{mau99}\\
${284  }$&{\bf 407} &${ 453 }$&{\bf 47.0} & ${+ 7.0}$ & ${- 8.0}$ & MSAM, \cite{msam}\\
${361  }$&{\bf 589} &${ 756 }$&{\bf 56.0} & ${+8.1 }$ & ${-6.9} $ & Ring5M2, \cite{lei98}\\
${543  }$&{\bf 615} &${ 717 }$&{\bf 49.0} & ${+19.1}$ & ${-13.6}$ & CAT2, \cite{sco96}  $^{*)}$\\[4pt]
\hline
\end{tabular}
\end{center} 
 $ ^{*)}$ - $\ell_{eff}$ and band width were taken from Max Tegmark's CMB data analysis center (\cite{teg})
\label{Cldat}
\end{table*}

\subsection{Other experimental constraints}

A constraint of the amplitude of the fluctuation power spectrum at
cluster scale can be derived from the cluster mass and the X-ray temperature
functions. It is usually formulated as a constraint for the density
fluctuation in a top-hat sphere of 8\Mpc radius, $\sigma_{8}$, which
can be calculated for a given initial power spectrum $P(k)$:
\be
\sigma_{8}^{2}={1\over
2\pi^{2}}\int_{0}^{\infty}k^{2}P(k)W^{2}(8{\rm Mpc}\;k/h)dk,
\label{si8}
\ee
where $W(x)=3(\sin x-x \cos x)/x^3$ is the Fourier transform of a
top-hat window function.
A recent optical determination of the mass function of nearby galaxy
clusters (\cite{gir98}) gives $\tilde
\sigma_{8}\tilde\Omega_m^{0.46-0.09\Omega_m}=0.60\pm 0.04$.
Several groups have found similar results using  different
methods and different data sets (for a comprehensive list of references
see \cite{borg99}).
To take into account the results from other authors
we have decided to use more conservative error bars:
\be
\tilde
\sigma_{8}\tilde\Omega_m^{0.46-0.09\Omega_m}=0.60\pm 0.08~.
\ee
>From the existence of three very massive clusters of galaxies observed
so far at $z>0.5$ a further constraint has been established by
\cite{bah98}
\be
\tilde \sigma_8\tilde\Omega_m^{\alpha}=0.8\pm 0.1\;,
\ee
where $\alpha =0.24$ if $\Omega_{\Lambda}=0$ and $\alpha =0.29$ if
$\Omega_{\Lambda}>0$ with $\Omega_{\Lambda}+\Omega_m=1.$ The
relation of this value to other tests will be analyzed too.

Another constraint on the amplitude of the linear power spectrum of
density fluctuations in our vicinity comes from the study of galaxy
bulk flows in spheres of large enough radius around our
position. Since these data may be influenced by the local super-cluster
(cosmic variance), we will use only the value of bulk motion - the
mean peculiar velocity of galaxies in the sphere of radius
$50h^{-1}$Mpc given by \cite{kol97},
\be
\tilde V_{50}=(375\pm 85) {\rm km/s.}
\ee
An essential constraint on the linear power spectrum of matter
clustering on small scales ($k\sim (2-40)h/$Mpc comes
from the Ly-$\alpha$ forest of absorption lines seen in quasar spectra
(\cite{gn98,cr98} and references therein).  Assuming that the
Ly-$\alpha$ forest is formed by discrete clouds with a physical size close
to the Jeans scale in the reionized inter-galactic medium at $z\sim 2-4$,
\cite{gn98} has obtained a constraint on the value of the
r.m.s. linear density fluctuations
\bea
 1.6<\tilde \sigma_{F}(z=3)<2.6~~(95\% \mbox{C.L.}) &&\\
 \mbox{ at }~ k_{F}\approx 34\Omega_m^{1/2}h/{\rm Mpc}~. \nonumber
\eea
Taking into account the
new data on quasar absorption lines, the effective equation of state
and the temperature of the inter-galactic medium at high redshift
were re-estimated
recently by \cite{ric99}. As result the value of Jeans scale at $z=3$ has
moved to $k_{F}\approx 38\Omega_m^{1/2}h/$Mpc (\cite{gn99}).

The procedure of
recovering the linear power spectrum from the Ly-$\alpha$ forest has been
elaborated by \cite{cr98}. Analyzing the absorption lines in a sample
of 19 QSO spectra they have obtained the following constraint on the
amplitude and slope of the linear power spectrum at $z=2.5$ and
$k_{p}=1.5\Om_m^{1/2}h/$Mpc,
\be
\tilde \Delta_{\rho}^2(k_p)\equiv k_p^3P(k_p)/2\pi^2=0.57\pm 0.26,
\ee
\be
\tilde n_p\equiv {\Delta \log\;P(k)\over \Delta \log\;k}\mid
_{k_p}=-2.25\pm 0.18,
\ee
(95\% CL). In addition to the power spectrum measurements we will use
the constraints on the value of the Hubble constant
\be
\tilde h=0.65\pm 0.15
\ee
 which is a compromise between measurements made by two groups:
\cite{tam97} and \cite{mad98}. We also employ nucleosynthesis
constraints on the baryon density of
\be
\widetilde{\Omega_bh^2} = 0.019\pm 0.0024 (95\% {\rm CL})
\ee
given by \cite{bur99}. An earlier value
of $\widetilde{\Omega_bh^2}=0.024\pm 0.006$ by \cite{tyt96} will be
used to analyze the influence of this assumption on the obtained
cosmological parameters.

\section{Testing the Method}

In order to test our method to determine cosmological
parameters for stability, we have constructed a mock sample of
observational data. We start with a set of cosmological parameters and
determine for them the ``observational'' data which would be measured
in case of faultless measurements with $1\sigma$ errors comparable to
the observational errors.  We then insert random sets of starting
parameters into the search program  and try to find the right model
which corresponds to the mock data. The method is stable if we can
recover our input cosmological model. Even starting very far away from
the true values, our method reveals as very stable and finds the
'true' model whenever possible (see Table~\ref{testtab}).

One of the main ingredients for the solution for our search problem
is a reasonably fast and accurate determination of the  transfer function
which depends on the cosmological parameters. We use the accurate analytical
approximations of the MDM transfer function $T(k;z)$ depending on
the parameters $\Omega_m$, $\Omega_b$, $\Omega_{\nu}$, $N_{\nu}$ and
$h$ by (\cite{eh3} and \cite{nov98}).

The linear power spectrum of matter density fluctuations is
\be
P(k;z)=Ak^nT^2(k;z)D_1^2(z)/D_1^2(0),\label{pkz}
\ee
where $A$ is the normalization constant and $D_1(z)$ is the linear
growth factor, which can be approximated by (\cite{car92})
$$D_1(z)=
{5\over 2}{\Omega_m(z)\over 1+z}\left[{1\over
70}+{209\Omega_m(z)-\Omega_m^2(z) \over 140}+\Omega_m^{4/7}(z)\right]^{-1},$$
where
$\Omega_m(z)=\Omega_m(1+z)^3/\left(\Omega_m(1+z)^3+\Omega_{\Lambda}\right)$.

  We normalize the spectra to the 4-year COBE data
which determines the amplitude of density perturbation at the horizon
crossing scale, $\delta_h$ (\cite{lid96,bun97}), which for a matter dominated
Universe without tensor mode and cosmological constant is given by
\be
\delta_h=1.95\times 10^{-5}\Omega_m^{-0.35-0.19\ln{\Omega_m}-0.17\tilde n}e^{
-\tilde n -0.14\tilde n^2}\label{dh1}~.
\ee
 For a flat model with cosmological constant
($\Omega_m+\Omega_{\Lambda}=1$) we have
\be
\delta_h=1.94\times 10^{-5}\Omega_m^{-0.785-0.05\ln{\Omega_m}}e^{
-0.95\tilde n -0.169\tilde n^2}\label{dh2}
\ee
($\tilde n\equiv n-1$). The normalization constant $A$ is then given by
\be
A=2\pi^{2}\delta_{h}^{2}(3000/h)^{3+n}
\;{\rm Mpc}^{4}.\label{anorm}
\ee

The Abell-ACO power spectrum is related to the matter power
spectrum at $z=0$, $P(k;0)$ by the cluster biasing parameter $b_{cl}$.
We assume scale-independent, linear bias:
\be
P_{A+ACO}(k)=b_{cl}^{2}P(k;0).\label{pcl}
\ee
For a given set of parameters $n$, $\Omega_m$, $\Omega_b$, $h$,
$\Omega_{\nu}$, $N_{\nu}$ and $b_{cl}$ theoretical values of
$P_{A+ACO}(k_j)$ can now be obtained for the values $k_j$ of Table 1. We
denote them by $y_j$ ($j=1,...,13$).

\begin{figure}[th]
\epsfxsize=9truecm
\epsfbox{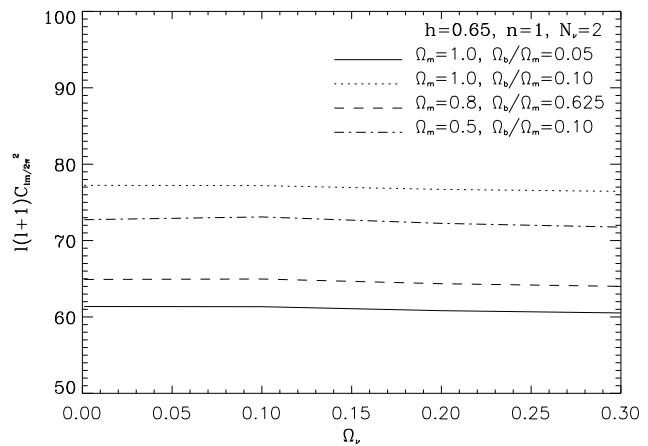}
\caption{The  dependence of the acoustic peak amplitude $A_p$ on neutrino
content $\Omega_{\nu}$ }
\label{dpa}
\end{figure}

\begin{figure}[th]
\epsfxsize=9truecm
\epsfbox{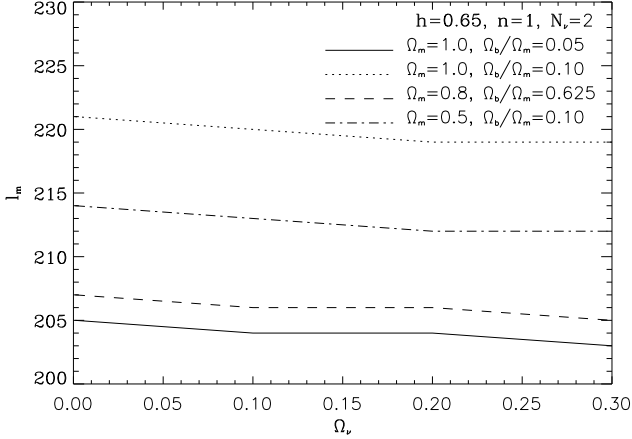}
\caption{The dependence of the acoustic peak position $\ell_p$ on neutrino
content $\Omega_{\nu}$ }
\label{dpl}
\end{figure}

The dependence of the position and amplitude of the first acoustic
peak of the CMB power spectrum on cosmological
parameters has been investigated using CMBfast by
\cite{sz96}.  As expected, the results are, within sensible
accuracy, independent of the hot dark matter
contribution ($\Om_\nu$).
This is illustrated in Figs.~\ref{dpa} and \ref{dpl}.
For the remaining parameters, $n$, $h$, $\Omega_b$ and
$\Omega_{\Lambda}$, we have determined the resulting values $\ell_p$ and
$A_p$ with CMBfast for a network of model parameters. The values
$\ell_p,A_p$ in-between grid points are then obtained by 4-dimensional
interpolation. This allows a fast and sufficiently accurate
calculation of the peak position and amplitude for a given set of parameters
in the range $ 0.7 \le n \le 1.4$, $0.3 \le h \le 0.8$,
$0 \le \Omega_b \le 0.2 $ and $0 \le \Omega_{\Lambda} \le 0.8$
considered in this work.  The
accuracy of this interpolation is estimated to be within 2\%. We
denote $\ell_p$ and $A_p$ by $y_{14}$ and $y_{15}$ respectively.

The theoretical values of the other experimental constraints are
obtained as follows: The density fluctuation $\sigma_8$ is
calculated according to Eq.~(\ref{si8}) with $P(k;z)$ taken from
Eq.~(\ref{pkz}). We
set $y_{16}=\sigma_{8}\Omega_m^{0.46-0.09\Omega_m}$ and $y_{17} =
\sigma_{8}\Omega^{\alpha}$, where $\alpha =0.24$ for
$\Omega_{\Lambda}=0$ and $\alpha =0.29$ for $\Omega_{\Lambda}>0$,
respectively.

The r.m.s. peculiar velocity of
galaxies in a sphere of radius $R=50h^{-1}$Mpc is given by
\be
V^{2}_{50}={1\over 2\pi^{2}}
\int_{0}^{\infty}
k^2P^{(v)}(k)e^{-k^{2}R_{f}^{2}}W^{2}(50{\rm Mpc}~k/h)dk,  \label{V50th}
\ee
where $P^{(v)}(k)$ is power spectrum for the velocity field of the
density-weighted matter (\cite{eh3}),
$W(50{\rm Mpc}~k/h)$ is the top-hat window function.
A previous smoothing
of raw data with a Gaussian filter of radius $R_{f}=12h^{-1}$Mpc is
employed here similar to the procedure which has led to the observational
value.  For the scales of interest
 $P^{(v)}(k)\approx (\Omega^{0.6}H_0)^2P(k;0)/k^2$.
We denote the r.m.s. peculiar velocity by $y_{18}$.

The  value by \cite{gn98} from the formation of Ly-$\alpha$ clouds
constrains the r.m.s. linear density perturbation at $z=3$ and
$k_{F}=38\Omega_m^{1/2}h/$Mpc. In terms of the power spectrum $\sigma_F$
is given by
\be
\sigma_{F}^{2}(z)={1\over
2\pi^{2}}\int_{0}^{\infty}k^{2}P(k;z)e^{(-k/k_F)^2}dk,\label{siF}.
\ee
It will be denoted by $y_{19}$. The corresponding value of the
constraint by \cite{cr98} is
\be
\Delta_{\rho}^2(k_p,z)\equiv k_p^3P(k_p,z)/2\pi^2,   \label{Dekp}
\ee
at $z=2.5$ and $k_{p}=0.008H(z)/(1+z)({\rm km/s})^{-1}$,
(where $H(z)=H_0\left[\Omega_m(1+z)^3+\Omega_{\Lambda}\right]^{1/2}$ is the
Hubble parameter at redshift z) will be denoted by $y_{20}$. The slope
of the power spectrum at this scale and redshift,
\be
n(z)\equiv {\Delta \log\;P(k,z)\over \Delta \log\;k}~,\label{enp}
\ee
is denoted by $y_{21}$.

For all tests except Gnedin's Ly-$\alpha$ clouds
 we used the density weighted  transfer function $T_{cb\nu}(k,z)$
from \cite{eh3}. For Gnedin's $\sigma_F$ we use $T_{cb}(k,z)$
according to the prescription of (\cite{gn98}). It must be noted that
even in the model with maximal $\Omega_{\nu}$ ($\sim0.2$) the difference
between $T_{cb}(k,z)$ and $T_{cb\nu}(k,z)$ is less than $12\%$ for  $k\le k_p$.

Finally, the values $\Omega_b$
and $h$ are denoted by $y_{22}$ and $y_{23}$ respectively.

The relative quadratic deviations of the
theoretical values from their observational counterparts are given
 by $\chi^2$:
\be
\chi^{2}=\sum_{j=1}^{23}\left({\tilde y_j-y_j \over \Delta \tilde y_j}
\right)^2,     \label{chi2}
\ee
where $\tilde y_j$ and $\Delta \tilde y_j$ are the experimental data
and their dispersion, respectively. The set of parameters $n$,
$\Omega_m$, $\Omega_b$, $h$, $\Omega_{\nu}$, $N_{\nu}$
and $b_{cl}$ or
some subset of them can be determined by minimizing $\chi^2$ using the
Levenberg-Marquardt method (\cite{nr92}). The derivatives of the predicted
values with respect to the search parameters which are required by this method
are calculated numerically using a relative step size of $10^{-5}$ with respect
to the given parameter.
\begin{table*}[th]
\caption{{\bf Test of the method:} results of  parameter search from
mock data for the tilted $\Lambda$MDM model ($n=1.2$,
$\Omega_m=0.55$, $\Omega_b=0.06$, $\Omega_{\nu}=0.20$, $N_{\nu}=2$,
$h=0.65$). In test~1, all parameters are determined; in the 2nd to 6th tests,
some parameters are fixed. For each test the first row corresponds to
the case when number of species of massive neutrinos is equal the input
value (2) and the second - when $N_{\nu}=3$.}
\begin{center}
\def\onerule{\noalign{\medskip\hrule\medskip}}
\medskip
\begin{tabular}{|ccccccccc|}
\hline
 & & & & & & & & \\
No   &$N_{\nu}$    & $\chi^2_{min}$  &$n$ &$\Omega_m$  &
 $\Omega_{\nu}$& $\Omega_b$ & $h$   & $b_{cl}$ \\ [4pt]
\hline
 & & & & & & & & \\
1 &2& 0.00&1.20$\pm$0.07&0.55$\pm$0.15&0.200$\pm$0.059&0.060$\pm$0.022&0.65$\pm$0.12&3.00$\pm$0.45  \\
  &3& 0.05&1.21$\pm$0.07&0.63$\pm$0.17&0.251$\pm$0.073&0.061$\pm$0.022&0.65$\pm$0.12&3.12$\pm$0.46  \\ [4pt]
2 &2& 1.72&1.18$\pm$0.06&0.79$\pm$0.07&0.281$\pm$0.055&0.101$\pm$0.005&0.50$ ^{*)}$&3.58$\pm$0.32    \\
  &3& 3.84&1.21$\pm$0.06&0.98$\pm$0.08&0.446$\pm$0.035&0.101$\pm$0.005&0.50$ ^{*)}$&3.65$\pm$0.30     \\ [4pt]
3 &2& 0.00&1.20$\pm$0.07&0.55$\pm$0.05&0.200$\pm$0.036&0.060$\pm$0.003&0.65$ ^{*)}$&3.00$\pm$0.27     \\
  &3& 0.05&1.21$\pm$0.07&0.62$\pm$0.05&0.249$\pm$0.043&0.060$\pm$0.003&0.65$ ^{*)}$&3.10$\pm$0.27    \\ [4pt]
4 &2& 0.68&1.21$\pm$0.07&0.45$\pm$0.04&0.168$\pm$0.029&0.045$\pm$0.002&0.75$ ^{*)}$&2.73$\pm$0.24 	  \\
  &3& 0.80&1.22$\pm$0.07&0.51$\pm$0.05&0.207$\pm$0.034&0.045$\pm$0.002&0.75$ ^{*)}$&2.83$\pm$0.25    \\ [4pt]
5 &2& 0.00&1.20$\pm$0.07&0.55$\pm$0.05&0.200$\pm$0.036&0.060$ ^{*)}$&0.65$ ^{*)}$&3.00$\pm$0.27      \\
  &3& 0.05&1.21$\pm$0.07&0.62$\pm$0.05&0.249$\pm$0.043&0.060$ ^{*)}$&0.65$ ^{*)}$&3.10$\pm$0.27       \\ [4pt]
6 &2&18.97&1.09$\pm$0.06&0.30$ ^{*)}$&0.039$\pm$0.003&0.047$\pm$0.007&0.73$\pm$0.05&3.60$\pm$0.38    \\
  &3&18.20&1.02$\pm$0.08&0.30$ ^{*)}$&0.000$\pm$0.001&0.064$\pm$0.021&0.63$\pm$0.10&3.55$\pm$0.31     \\ [4pt]
\hline
\end{tabular}
\end{center}
 $ ^{(*)}$ - fixed parameters.  \label{testtab}
\end{table*}

The method was tested in the following way. Assuming a 4-year COBE
normalized tilted $\Lambda$MDM model with the parameters $n=1.2$,
$\Omega_m=0.55$, $\Omega_b=0.06$, $\Omega_{\nu}=0.2$, $N_{\nu}=2$,
$h=0.65$ and assuming further a cluster biasing parameter
$b_{cl}=3.0$ we have calculated mock cluster power spectrum $\tilde
P_{A+ACO}(k_j)$ and treated them as $\tilde y_{i}$, $i=1,...,13$.  The
remaining mock data $\tilde y_{i}$, $i=14,...,23$ have been calculated
as described above.  We have assigned to these mock data the same
relative 'experimental' errors as in the corresponding experiments 
described in the previous section.

We then used these mock data to search the parameters $n$,
$\Omega_m$, $\Omega_b$, $h$, $\Omega_{\nu}$, and $b_{cl}$ ($N_{\nu}$
was fixed). As starting parameters for the search program we assumed
random values within the allowed range.  We have searched for the
parameters assuming the ``true'' value of two species of massive
neutrinos as well as assuming three species of massive neutrinos.  The
parameters obtained for different cases are presented in Table 3.
The errors in the determined parameters are calculated as root square
from diagonal elements of the standard error covariance matrix. In
all cases the code found all the previous known parameters with high
accuracy. This means that the code finds the global minimum of
$\chi^2$ independent of the initial values for the parameters.

Our conclusions from the test results can be summarized as follows:

1. If all parameters are free and $N_{\nu} = 2 $ (the input
value) the code finds the correct values of the free parameters (test 1,
for $N_{\nu}=2$ in Table 3).

2. If all parameters are free and $N_{\nu} = 3 $
the code finds values of the free parameters which are in the
$1\sigma$ range of errors (test 1, for $N_{\nu}=3$ in Table 3).

3. If some parameters are fixed and differ from the input values
(tests 2, 4, 6 in Table 3) the code finds for the remaining search
parameters values close to the correct ones. The most stable and accurate
value is $\Omega_b$. The results for $n$,
$\Omega_{\nu}$ and $\Omega_m$ are in the $\le 2\sigma$ range of the
correct values. The most uncertain solutions are found for $n$ and
$\Omega_{\nu}$ if an incorrect value for $\Omega_m$ has been
assumed (test 6 in Table 3).

4. If some parameters are fixed to the predetermined ones and $N_{\nu}
= 2 $ (the input value) the code finds the correct values of the free
parameters (test 3 and 5 , for $N_{\nu}=2$ in Table 3), if $N_{\nu} =
3 $ the determined values are within the $1\sigma$ range (test 3 and
5 , for $N_{\nu}=3$ in Table 3).

In summary, the code determines the parameters
$n$, $\Omega_{\nu}$, $\Omega_b$, $h$, $b_{cl}$ and $\Omega_m$
correctly, if the observational data are correctly measured and the
 cosmological model assumed is correct; {\em i.e.} no curvature,
a negligible amount of tensor perturbations and a
primordial spectrum of scalar perturbations which is scale free
from the present horizon size down to the scale of the Ly-${\alpha}$ clouds.

\section{Results}

The determination of the parameters $n$, $\Omega_m$, $\Omega_b$, $h$,
$\Omega_{\nu}$, $N_{\nu}$ and $b_{cl}$ by the Levenberg-Marquardt $\chi^2$
minimization method can be realized in the following way: We vary the
set of parameters $n$, $\Omega_m$, $\Omega_b$, $h$,
$\Omega_{\nu}$  and $b_{cl}$ or some subset of them and find the
minimum of
$\chi^2$. Since the $N_{\nu}$ process is discrete we repeat this
procedure three times for $N_{\nu}$=1, 2, and 3.  The lowest of the
three minimums is the minimum of $\chi^2$ for the
complete set of free parameters. The number of degrees of
freedom $N_F= N_{\rm exp}-N_{\rm par}= 7$ if all parameters are
free. It increases, if some of the parameters are fixed to a certain value.
(Remember that even though we have 13 power spectra points, they can
be described by just 3 degrees of freedom.)

We have determined the minimum of $\chi^2$ for $N_{\nu}$=1, 2, 3 in
11 different cases, where all observational data described in Sect. 2
are used.

1) $n$, $\Omega_m$, $\Omega_{\nu}$, $\Omega_b$, $h$, and $b_{cl}$ are
   free parameters ($N_F=7$);

2) $h=0.5$ is fixed, the remaining parameters are free ($N_F=8$);

3) $h=0.6$ (\cite{Saha99},\cite{Tamm99}) is fixed, the remaining
   parameters are free ($N_F=8$);

4) $h=0.72$ (\cite{mad98,Richtler99}) is fixed, the remaining
   parameters are free ($N_F=8$);

5) $h=0.6$ (\cite{Saha99},\cite{Tamm99}) and  $h^2\Omega_b=0.024$
   (\cite{tyt96}) are fixed, the remaining parameters are free ($N_F=9$);

6) $\Omega_m=1.0$ is fixed, the remaining parameters are free ($N_F=8$);

7) $\Omega_m=0.3$ is fixed, the remaining parameters are free  ($N_F=8$);

8) $n=1$ is fixed, the remaining parameters are free ($N_F=8$);

9) $n=1$, $\Omega_m=1$ are fixed, the remaining parameters are free
($N_F=9$);

10) $n=1$, $\Omega_m=0.3$ and are fixed, the remaining parameters are
free ($N_F=9$);

11) $\Omega_{\nu} =7.4\times10^{-4}N_{\nu}/h^2$ is fixed by the lower limit of neutrino
    mass inferred by the observed neutrino oscillations in the
    Super-Kamiokande experiment ($N_F=8$).

For these 11 cases we find the  minimum of $\chi^2$ from
which we determine the parameters presented in Table 4. Note, that
for all models $\chi^2_{min}$ is in the range,
$N_F-\sqrt{2N_F}\le \chi^2_{\min}\le N_F+\sqrt{2N_F}$ which is
expected for a Gaussian distribution of $N_F$
degrees of freedom. This means that the cosmological paradigm which
has been  assumed is in agreement with the
data.
(Note here, that the reduction of the 13 not
independent data points of the cluster power spectrum to three parameters is very
important for our analysis. Otherwise we would obtain a $\chi_{\min}$
which is by far too small.
If we would have assumed the 13 points of the Abell cluster
power spectrum as independent, resulting in $N_F=17$, the smallness of
$\chi_{\min}$ would have indicated that something is wrong in our
approach. But we might have drawn the wrong conclusion that the error
bars be too large!)
In Table 5 we present also the values of the different observational
constraints for the best fit models found in Table~4.

\begin{table*}[th]
\caption{Cosmological parameters determined for the tilted
$\Lambda$MDM model with one, two and three species of massive
neutrinos. In case No. 1 all parameters are free, in the other cases
(No2-11) some of them are fixed, as described above. }
\begin{center}
\def\onerule{\noalign{\medskip\hrule\medskip}}
\medskip
\begin{tabular}{|ccccccccc|}
\hline
&&&&&&&&\\
No  &$N_{\nu}$     & $\chi^2_{min}$  &$n$  & $\Omega_m$&$\Omega_{\nu}$& $\Omega_b$ & $h$    & $b_{cl}$ \\ [4pt]
\hline
&&&&&&&&\\
1&1 & 4.64&1.12$\pm$0.09&0.41$\pm$0.11&0.059$\pm$0.028&0.039$\pm$0.014&0.70$\pm$0.12&2.23$\pm$0.33\\
 &2 & 4.82&1.13$\pm$0.10&0.49$\pm$0.13&0.103$\pm$0.042&0.039$\pm$0.014&0.70$\pm$0.13&2.33$\pm$0.36\\
 &3 & 5.09&1.13$\pm$0.10&0.56$\pm$0.14&0.132$\pm$0.053&0.040$\pm$0.015&0.69$\pm$0.13&2.45$\pm$0.37\\ [4pt]
2&1 & 7.50&1.11$\pm$0.09&0.64$\pm$0.10&0.075$\pm$0.058&0.076$\pm$0.005&0.50$ ^{*)}$&2.72$\pm$0.28 \\
 &2 & 7.46&1.12$\pm$0.09&0.73$\pm$0.12&0.120$\pm$0.075&0.076$\pm$0.005&0.50$ ^{*)}$&2.86$\pm$0.28 \\
 &3 & 7.46&1.13$\pm$0.09&0.82$\pm$0.14&0.163$\pm$0.089&0.076$\pm$0.005&0.50$ ^{*)}$&2.96$\pm$0.29 \\ [4pt]
3&1 & 5.28&1.12$\pm$0.09&0.51$\pm$0.07&0.074$\pm$0.041&0.053$\pm$0.003&0.60$ ^{*)}$&2.43$\pm$0.26 \\
 &2 & 5.45&1.13$\pm$0.09&0.59$\pm$0.08&0.110$\pm$0.053&0.053$\pm$0.003&0.60$ ^{*)}$&2.56$\pm$0.26 \\
 &3 & 5.62&1.13$\pm$0.09&0.66$\pm$0.10&0.144$\pm$0.063&0.053$\pm$0.003&0.60$ ^{*)}$&2.66$\pm$0.27 \\ [4pt]
4&1 & 4.67&1.12$\pm$0.10&0.39$\pm$0.05&0.058$\pm$0.026&0.037$\pm$0.002&0.72$ ^{*)}$&2.19$\pm$0.23 \\
 &2 & 4.84&1.13$\pm$0.06&0.47$\pm$0.06&0.101$\pm$0.014&0.037$\pm$0.002&0.72$ ^{*)}$&2.29$\pm$0.18 \\
 &3 & 5.12&1.14$\pm$0.10&0.53$\pm$0.07&0.130$\pm$0.046&0.037$\pm$0.002&0.72$ ^{*)}$&2.38$\pm$0.25 \\ [4pt]
5&1 & 5.68&1.11$\pm$0.09&0.53$\pm$0.07&0.068$\pm$0.043&0.067$ ^{*)}$&0.60$ ^{*)}$&2.49$\pm$0.27   \\
 &2 & 5.76&1.11$\pm$0.09&0.61$\pm$0.09&0.103$\pm$0.056&0.067$ ^{*)}$&0.60$ ^{*)}$&2.62$\pm$0.27     \\
 &3 & 5.85&1.12$\pm$0.09&0.67$\pm$0.10&0.136$\pm$0.067&0.067$ ^{*)}$&0.60$ ^{*)}$&2.71$\pm$0.27     \\ [4pt]
6&1 &12.23&1.07$\pm$0.09&1.00$ ^{*)}$&0.116$\pm$0.086&0.118$\pm$0.027&0.40$\pm$0.05&3.15$\pm$0.39 \\
 &2 &10.18&1.10$\pm$0.09&1.00$ ^{*)}$&0.177$\pm$0.086&0.099$\pm$0.022&0.44$\pm$0.05&3.10$\pm$0.38 \\
 &3 & 8.80&1.12$\pm$0.09&1.00$ ^{*)}$&0.219$\pm$0.084&0.085$\pm$0.019&0.47$\pm$0.05&3.07$\pm$0.38 \\ [4pt]
7&1 & 6.55&1.04$\pm$0.10&0.30$ ^{*)}$&0.000$\pm$0.005&0.038$\pm$0.013&0.71$\pm$0.12&2.25$\pm$0.19 \\
 &2 & 6.55&1.04$\pm$0.10&0.30$ ^{*)}$&0.000$\pm$0.005&0.038$\pm$0.013&0.71$\pm$0.12&2.25$\pm$0.19 \\
 &3 & 6.55&1.04$\pm$0.10&0.30$ ^{*)}$&0.000$\pm$0.005&0.038$\pm$0.013&0.71$\pm$0.12&2.25$\pm$0.19 \\ [4pt]
8&1 & 6.21&1.00$ ^{*)}$&0.45$\pm$0.12&0.042$\pm$0.033&0.038$\pm$0.014&0.71$\pm$0.13&2.44$\pm$0.31 \\
 &2 & 6.60&1.00$ ^{*)}$&0.50$\pm$0.14&0.062$\pm$0.043&0.038$\pm$0.014&0.71$\pm$0.13&2.57$\pm$0.32 \\
 &3 & 6.85&1.00$ ^{*)}$&0.51$\pm$0.14&0.063$\pm$0.012&0.038$\pm$0.014&0.71$\pm$0.13&2.69$\pm$0.32 \\ [4pt]
9&1 &13.01&1.00$ ^{*)}$&1.00$ ^{*)}$&0.088$\pm$0.075&0.104$\pm$0.027&0.43$\pm$0.05&3.23$\pm$0.36  \\
 &2 &11.46&1.00$ ^{*)}$&1.00$ ^{*)}$&0.130$\pm$0.073&0.086$\pm$0.024&0.47$\pm$0.06&3.23$\pm$0.35  \\
 &3 &10.46&1.00$ ^{*)}$&1.00$ ^{*)}$&0.159$\pm$0.069&0.075$\pm$0.021&0.51$\pm$0.07&3.23$\pm$0.35  \\ [4pt]
10&1& 6.94&1.00$ ^{*)}$&0.30$ ^{*)}$&0.000$\pm$0.010&0.034$\pm$0.009&0.75$\pm$0.09&2.25$\pm$0.20  \\
  &2& 6.94&1.00$ ^{*)}$&0.30$ ^{*)}$&0.000$\pm$0.010&0.034$\pm$0.009&0.75$\pm$0.10&2.25$\pm$0.20  \\
  &3& 6.94&1.00$ ^{*)}$&0.30$ ^{*)}$&0.000$\pm$0.010&0.034$\pm$0.009&0.75$\pm$0.10&2.25$\pm$0.20  \\ [4pt]
11&1& 6.34&1.04$\pm$0.07&0.35$\pm$0.10&0.002$ ^{**)}$&0.045$\pm$0.016&0.65$\pm$0.11&2.36$\pm$0.26 \\
  &2& 6.48&1.04$\pm$0.07&0.36$\pm$0.10&0.004$ ^{**)}$&0.045$\pm$0.016&0.65$\pm$0.11&2.38$\pm$0.26  \\
  &3& 6.63&1.04$\pm$0.07&0.37$\pm$0.10&0.005$ ^{**)}$&0.046$\pm$0.016&0.64$\pm$0.11&2.41$\pm$0.26  \\ [6pt]
\hline
\end{tabular}
\end{center}
 $ ^{*)}$ - fixed parameters, $^{**)}$ mass density of neutrino is
fixed by the lowest limit of the neutrino mass from Super-Kamiokande results,
$h^2\Omega_{\nu}=\sqrt{\delta m^2}N_{\nu}/94eV$ with
$\sqrt{\delta m^2}=0.07$eV.
\end{table*}

\begin{table*}[th]
\caption{Theoretical  predictions for the observational values of
tilted $\Lambda$MDM  models found with the parameters of Table 4 (for the
value of $N_\nu$ leading to the lowest $\chi^2$).}
\begin{center}
\def\onerule{\noalign{\medskip\hrule\medskip}}
\medskip
\begin{tabular}{|ccccccccccc|}
\hline
&&&&&&&&&&\\
No&$N_{\nu}$& $\ell_p$ & $A_p$&$\sigma_8\Omega_{m}^{0.46-0.09\Omega_m}$&$\sigma_8\Omega_{m} ^{0.29}$&$V_{50},$km/s&$\sigma_{F}$&
$\Delta^2_{\rho}(k_p)$&$n_{p}(k_p)$&$t_0/10^9$yrs\\ [6pt]
\hline
&&&&&&&&&&\\
1& 1&215& 84.1& 0.65& 0.74& 353& 2.03& 0.48&-2.28&12.3 \\[4pt]
2& 1&233& 91.6& 0.68& 0.71& 348& 1.87& 0.51&-2.16&15.0 \\[4pt]
3& 1&223& 87.8& 0.67& 0.73& 356& 1.91& 0.52&-2.18&13.5 \\[4pt]
4& 1&214& 83.2& 0.65& 0.74& 353& 2.04& 0.49&-2.28&12.2 \\[4pt]
5& 1&224& 89.3& 0.67& 0.72& 353& 1.89& 0.52&-2.18&13.3 \\[4pt]
6& 3&228& 84.7& 0.72& 0.72& 366& 1.78& 0.44&-2.16&13.9 \\[4pt]
7& 1&217& 80.7& 0.59& 0.70& 297& 2.08& 0.68&-2.17&13.4 \\[4pt]
8& 1&209& 69.5& 0.65& 0.72& 323& 1.88& 0.58&-2.22&11.9 \\[4pt]
9& 3&218& 68.9& 0.71& 0.71& 331& 1.72& 0.47&-2.21&12.8 \\[4pt]
10&1&213& 73.4& 0.59& 0.70& 292& 2.00& 0.67&-2.20&12.5\\ [4pt]
11&1&221& 80.9& 0.62& 0.71& 300& 2.03& 0.65&-2.18&14.1 \\ [4pt]
Obs.& data&$253\pm 70$&$80\pm 17$&$0.60\pm 0.08$&$0.8\pm 0.1 $&$375\pm  85$&$2.0\pm .3$&$0.57\pm 0.26$&$-2.25\pm
0.2$&$13.2\pm 3$\\ [6pt] \hline
\end{tabular}
\end{center}
\end{table*}

\begin{figure}[th]
\epsfxsize=9truecm
\epsfbox{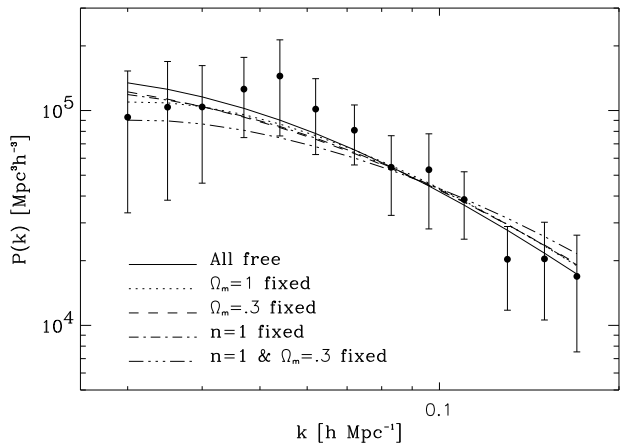}
\caption{The observed Abell-ACO power spectrum (filled circles) and
the theoretical spectra predicted by tilted $\Lambda$MDM models with
parameters taken from Table 4 ($N_{\nu}=1$).}
\label{pkth}
\end{figure}

If all parameters are free (Table 4, case No 1), the
model with one sort of massive neutrinos provides the best fit to
the data, $\chi^2_{min}\approx 4.6$. Note, however, that there are
only marginal differences in $\chi^2_{min}$ for $N_\nu =
1,2,3$. Therefore, with the given accuracy of the data we cannot conclude
whether -- if massive neutrinos are present
at all -- their number is one, two, or three. We summarize, that the
considered observational data on LSS of the Universe can be
explained by a flat $\Lambda$MDM inflationary model with a tilted
spectrum of scalar perturbations and vanishing tensor contribution.
The best fit parameters are: $n=1.12\pm0.10$, $\Omega_m=0.41\pm0.11$,
$\Omega_{\nu}=0.059\pm0.028$, $N_{\nu}=1$, $\Omega_b=0.039\pm0.014$ and $h=0.70\pm0.12$.
The CDM density parameter is $\Omega_{cdm} = 0.31\pm0.15$ and
$\Omega_{\Lambda}$  is considerable, $\Omega_{\Lambda}=0.59\pm0.11$.

The value of the Hubble constant is close to measurements by
\cite{mad98}. The spectral index coincides with
the COBE prediction.
The neutrino matter density $\Omega_{\nu}=0.059\pm0.028$
corresponds to a neutrino mass $m_{\nu}=94\Omega_{\nu}h^2\approx2.7\pm1.2$ eV. The
estimated cluster bias parameter $b_{cl}=2.23\pm0.33$ fixes the amplitude
of the Abell-ACO power spectrum (Fig.~\ref{pkth}).  All  predictions
of the measurements summarized in Table 5 are close to the
experimental values and within the  error bars of the data.

The predicted position of the acoustic peak ($\ell_p=215$) is
systematically lower than the experimental value determined here
from the complete data set on $\Delta T/T$ ($\tilde \ell_p=253\pm
70$). This position is nearly fixed by the requirement
$\Omega_m+\Omega_\Lambda =1$ and is only weakly dependent of the
parameters varied in this study.
The  acoustic peak inferred by the Boomerang experiment (\cite{mau99})
is situated at $\ell\sim 200$ and prefers models which are very 
close to flat (\cite{mel99b}).
The  models with low $\Omega_m \sim 0.3$
(case No. 7 in Table 4) fit the observable data somewhat less good than the
best model ($\Delta\chi^2_{min}\approx2.0$)
but all predictions are still within the $1\sigma$ range.
These models prefer a high Hubble parameter,
$h\approx 0.7$ and no massive neutrinos, $\Om_\nu=0$.
On the contrary, the matter dominated
tilted MDM model ($\Omega_m=1$, models 6 in Table 4) prefers high
$\Omega_{\nu}=0.22$, three sort of massive neutrino and a 
 low  Hubble parameter, $h=0.47$.
This can be understood by considering one of the most serious problems
of standard CDM, namely that the model, when normalized to COBE, has
too much power on small scales. This problem can be solved either by
introducing HDM and thereby damping the spectrum on small scales or by
introducing a cosmological constant which leads mainly to a 'shift of the
power spectrum to the left'.

Another interesting correlation can be seen in Table 4, cases No
2-4, where we have fixed $h$. An increasing Hubble constant is
compensated by a
decreasing matter density, $\Omega_m$, ({\em i.e.} increasing cosmological
constant) and a decreasing baryon content due to the tight
nucleosynthesis constraint on $\Om_bh^2$.
Furthermore, increasing the number of massive neutrino species
$N_{\nu}$ from 1 to 3 leads to an increase of $\Omega_{\nu}$ from
0.06 to 0.13 and to a decrease of $\Omega_{\Lambda}$ from 0.59 to 0.43
(case 1).

If we use the nucleosynthesis constraint by  \cite{tyt96} (case No 5),
$\chi^2_{min}$ is slightly higher than in case No 3.

Now let us discuss models with a perfectly scale invariant primordial power
spectrum as predicted by the first inflationary models, $n=1$ fixed
(cases 8-10 in Tables 4). If all of the remaining parameters are
free (case 8) then this data set prefers a $\Lambda$MDM model with
parameters $\Omega_m=0.45\pm0.12$ and $h=0.71\pm0.13$ and a somewhat
lower neutrino content than the best fit model.  Models with low
matter content, $\Omega_m=0.3$, prefer a high Hubble parameter,
$h\simeq 0.75$ and no hot dark matter, $\Om_\nu=0$ (case No 10).
The matter dominated model
$\Omega_m=1$ (case No 9) is the standard MDM model with
$\Omega_{\nu}=0.16\pm0.07$, three sort of massive neutrino
($m_{\nu}=1.3\pm0.7$eV) and $h=0.51\pm0.07$.

If the HDM component is eliminated or $\Omega_{\nu}$ is fixed at the
small value defined by the lower limit of the neutrino mass
$\sqrt{\delta m_{\nu}^2}=0.07$ from the Super-Kamiokande
experiment $\Omega_{\nu}=7.4\times 10^{-4}N_{\nu}/h^2$, we obtain the
best-fit value for the matter density parameter
$\Omega_m\approx 0.39\pm0.11$ and Hubble constant $h=0.62\pm0.12$ (case No 11).

The experimental Abell-ACO power spectrum and the theoretical
predictions for some best fit models are shown in Fig.~\ref{pkth}.
  Recently it was shown
(\cite{nov99}) that due to the large error bars, the position of the peak of
$\tilde P(k)$ at $k\approx 0.05$h/Mpc does not  influence
the determination of the cosmological parameters significantly.
 Mainly the slope of the power spectrum on scales smaller than the
scale of the peak position determines the cosmological parameters.

\begin{figure*}[tp]
\epsfxsize=16truecm
\epsfbox{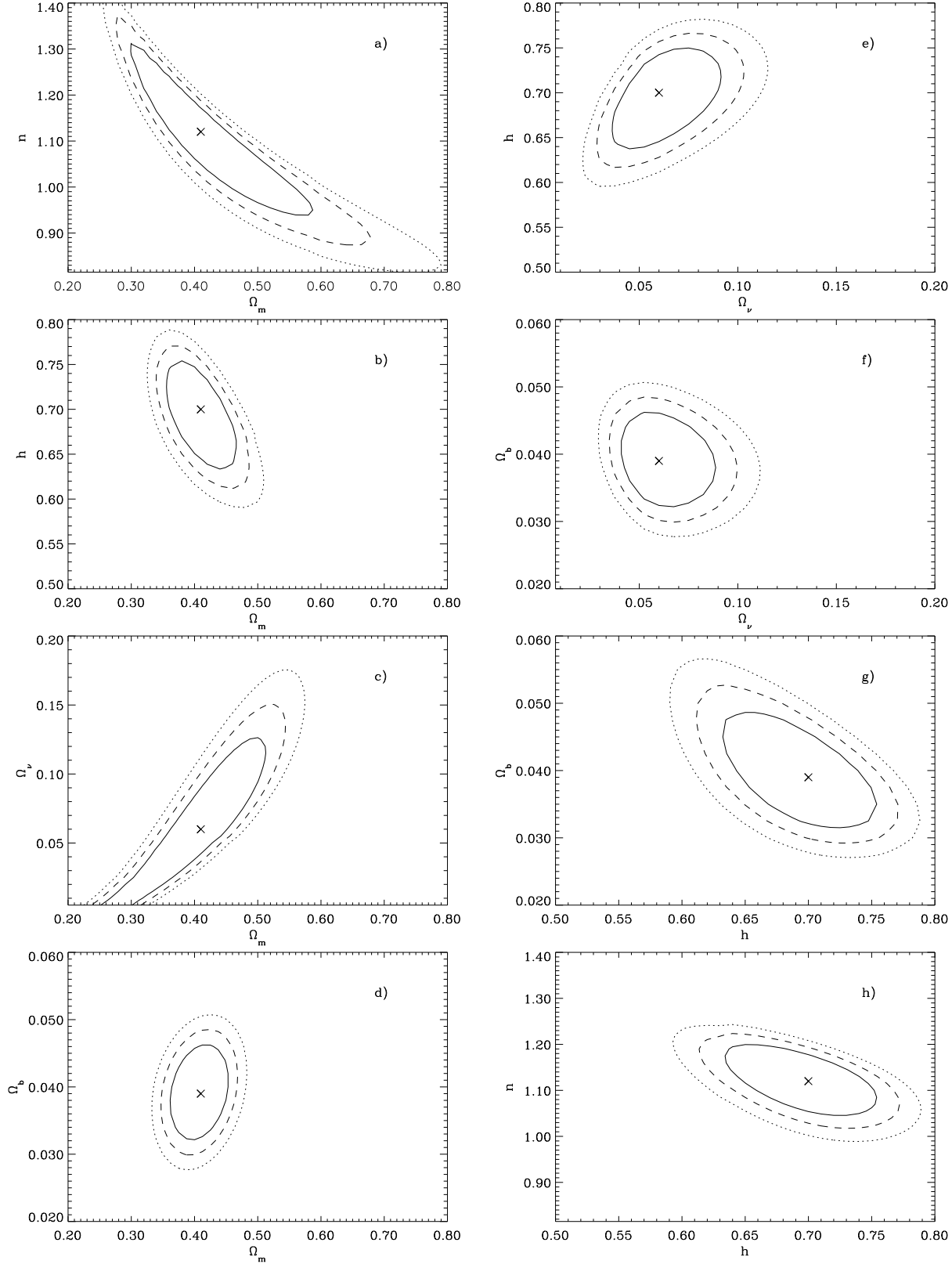}
\caption{Likelihood contours (solid line - 68.3\%, dashed - 95.4\%, dotted
- 99.73\%) of the tilted $\Lambda$MDM model with $N_{\nu}=1$ and
parameters from Table 4 (case 1) in the different planes of
$n-\Omega_m-\Omega_{\nu}-\Omega_b-h$ space.  The parameters
not shown in a given diagram are set to their best fit value.}
\label{Lcm1}
\end{figure*}

\begin{figure}[tp]
\epsfxsize=8truecm
\epsfbox{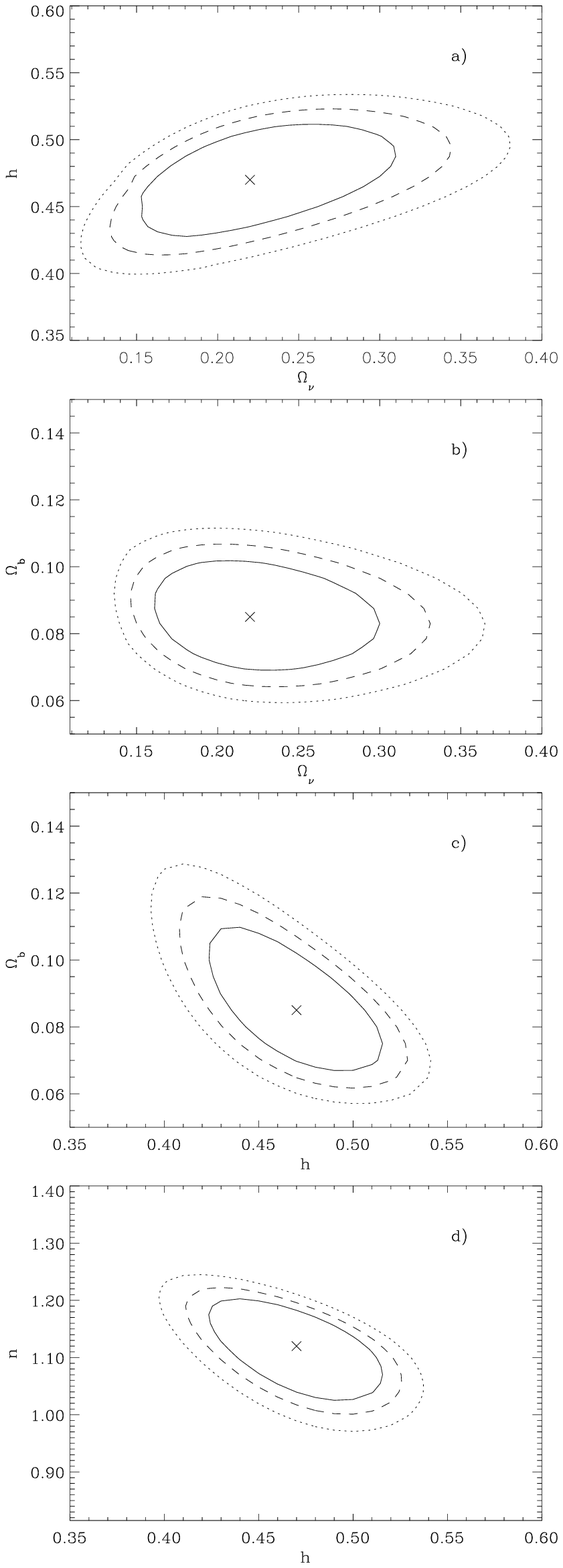}
\caption{Likelihood contours (solid line - 68.3\%, dashed - 95.4\%, dotted
- 99.73\%) of tilted $\Lambda$MDM with $N_{\nu}=3$, fixed
$\Omega_m=1$ and  parameters from Table 4 (case 6) in the
different planes of $n-\Omega_{\nu}-\Omega_b-h$ space. The parameters
not shown in  a given diagram are set to their best fit value.}
\label{Lcm6}
\end{figure}

\begin{figure}[tp]
\epsfxsize=8truecm
\epsfbox{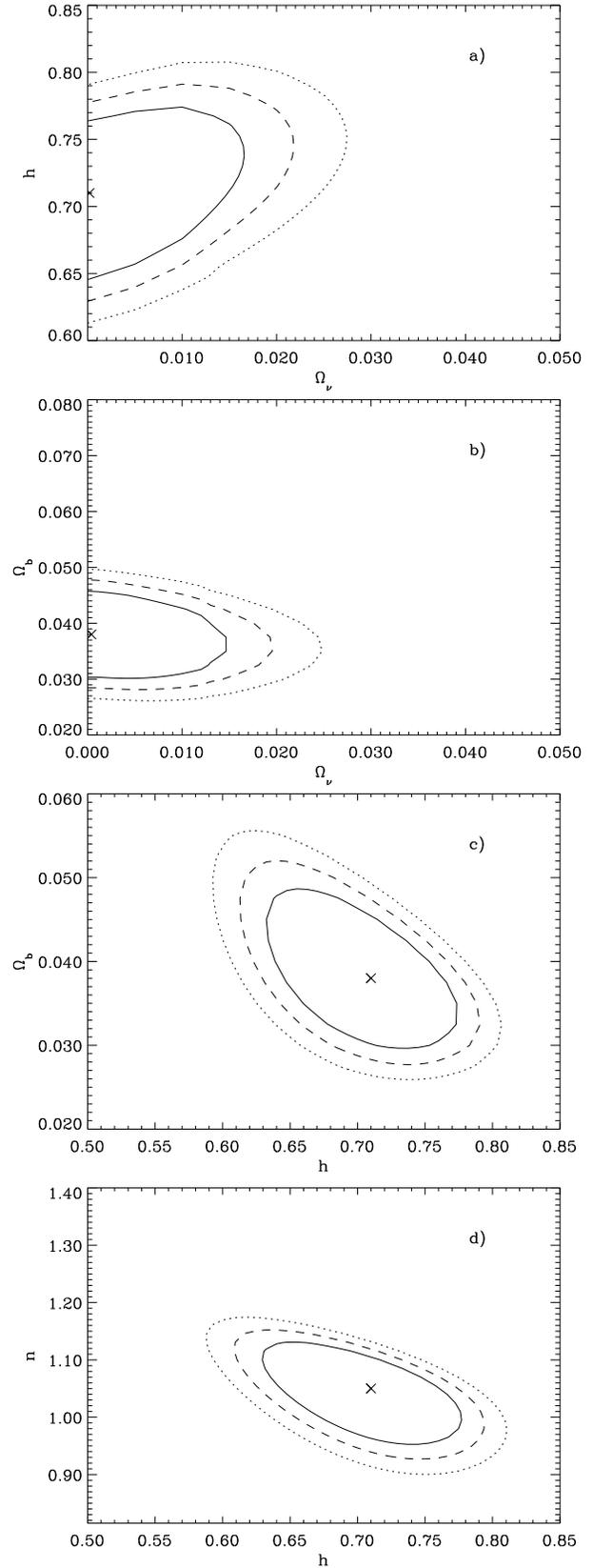}
\caption{Likelihood contours (solid line - 68.3\%, dashed - 95.4\%, dotted
- 99.73\%) of tilted $\Lambda$MDM with $N_{\nu}=3$, fixed
$\Omega_{m}=0.3$ and  parameters from Table 4 (case 7) in
the different planes of $n-\Omega_{\nu}-\Omega_b-h$ space.  The parameters
not shown in  a given diagram are set to their best fit value.}
\label{Lcm7}
\end{figure}

The errors in the best fit parameters presented in Table 4 are
the square roots of the  diagonal elements of the covariance
matrix.  More informations about the
accuracy of the determination of parameters and their sensitivity to
the data used can be obtained from the contours of confidence
levels presented in Fig.~\ref{Lcm1} for the tilted $\Lambda$MDM model with
parameters from Table 4 (case No 1, $N_{\nu}=1$).  The same contours
for cases No 6 and 7 are shown in Fig.~\ref{Lcm6} and \ref{Lcm7}, respectively.  These
contours show the confidence regions which contain 68.3\% (solid line),
95.4\% (dashed line) and 99.73\% (dotted line) of the total probability
distribution in the two dimensional sections of the six-dimensional
parameter space, if the probability distribution is Gaussian.
Since the number of degrees of freedom is
7 they correspond to $\Delta\chi^2=$8.2, 14.3 and 21.8
respectively. The parameters
not shown in a given diagram are set to their best-fit value.

As one can see in Fig.7a  the iso-$\chi^2$ surface is rather prolate from
the low-$\Omega_m$ - high-$n$  corner to high-$\Omega_m$ - low-$n$.
This indicates some degeneracy in $n-\Omega_m$ parameter plane, which
can be expressed by the following equation which roughly describes the
'maximum likelihood ridge' in this plane within the $1\si$:
\be
n\sqrt{\Omega_m}=0.73~.
\ee
A similar degeneracy is observed in the $\Omega_{\nu}- \Omega_m$ plane
in the range $0\le\Omega_{\nu}\le 0.17$, $0.25\le\Omega_m \le 0.6$
(Fig.7c). The equation for the 'maximum likelihood ridge'
or 'degeneracy equation' has here the form:
\be
\Omega_{\nu}=0.023-0.44\Omega_m+1.3\Omega_m^2~.
\ee

The 3rd column of Table 4 ($\chi^2_{min}$) shows that all models except 9th
with $N_{\nu}=1$ are within the 1$\sigma$ contour of the best fit. 

The next
important question is: which is the confidence limit of each parameter
marginalized over the other ones. The straight forward answer is the
integral of the likelihood function over the allowed range of all
the  other parameters.
But for a 6-dimensional parameter space this is computationally
time consuming.
Therefore, we have estimated the 1$\sigma$ confidence limits for
all parameters in
the following way. By variation of all parameter we determine
 the 6-dimensional $\chi^2$ surface which
contains  68.3\% of the total probability distribution.
We then project the surface onto  each axis
of parameter space. Its shadow on the parameter axes gives us the 1$\sigma$
confidence limits on cosmological parameters. For the best
$\Lambda$MDM model with one  sort of massive neutrinos the 1$\sigma$
confidence limits on parameters obtained  in this way are presented
in Table~\ref{tabmax}.
\begin{table}
\caption{\label{tabmax} The best fit values of all the parameters with
errors obtain by maximizing the (Gaussian) 68\% confidence contours
over all other parameters.}
\begin{center}
\begin{tabular}{||c|c||}
\hline
&\\
parameter & central value and errors\\ [4pt]
\hline
&\\
 $\Omega_m$ & $0.41^{+0.59}_{-0.22}$ \\[4pt]
$\Omega_{\nu}$ & $0.06^{+0.11}_{-0.06}$ \\[4pt]
 $\Omega_b$ & $0.039^{+0.09}_{-0.018}$ \\[4pt]
$h^{*)}$ & $0.70^{+0.15(+0.31)}_{-0.32}$ \\[4pt]
$n$ & $1.12^{+0.27}_{-0.30}$ \\[4pt]
 $b_{cl}$ & $2.22^{+1.3}_{-0.7}$\\[6pt]
\hline
\end{tabular}
\end{center}
 $ ^{*)}$ - the upper limit is obtained by including the
lower limit on the age of the Universe due to the age of oldest stars,
$t_0\ge13.2\pm 3.0$ (\cite{car99}). The value obtained without this
constraint is given in parenthesis.
\end{table}

It must be noted that the upper $1\sigma$ edge for $h$ is equal 1.08 when we marginalized
over all other parameters and input observable data used here. But
this  contradicts
the age of the oldest globular clusters $t_0=13.2\pm 3.0$ (\cite{car99}). Thus
we have included this value into the marginalization procedure for
the upper limit of $h$.  We then have 8 degrees of freedom (24 data points)
and the 6-dimensional $\chi^2$ surface which contains
68.3\% of the probability is confined by the value 13.95.
We did not use the age of oldest
globular cluster  for searching of best fit parameters in general case because it is
only a lower limit for age of the Universe, besides it does not change their values as
one can see from last column of Table 5.

The errors given  in Table~\ref{tabmax}  represent 68\% likelihood, of course,
only when the probability distribution is Gaussian. As one can see
from  Fig.7 (all panels without degeneracy) the ellipticity of the likelihood
contours in most of planes is close to what is expected from a Gaussian
distribution. This indicates that our estimates of the confidence limits
are reasonable.
These errors define the range of each parameter within which 
the best-fit values obtained for the remaining parameters lead to
$\chi^2_{min}\le 12.84$. Of course, the best-fit values of the 
remaining parameters lay within their corresponding
68\% likelihood given in the Table~\ref{tabmax}. It does however not mean
that any set of parameters from these ranges satisfies  the condition,
$\chi^2_{min}\le 12.84$.

 For example,
standard CDM model ($\Omega_m=1$, $h=0.5$, $\Omega_b=0.05$, $n=1$ and best-fit value of
cluster biasing parameter $b_{cl}=2.17$ ($\sigma_8=1.2$)) has $\chi^2_{min}=142$ (!),
that excludes it at very high confidence level, $>99.999\%$. When we use the
baryons density inferred from nucleosynthesis 
($h^2\Omega_b=0.019$ ($b_{cl}=2.25$, $\sigma_8=1.14$))
the situation does not improve much, $\chi^2_{min}=112$. Furthermore,
even if  we leave $h$ as free parameter we still find $\chi^2_{min}=16$ 
($>1\sigma$) with the best-fit values  $h=0.37$ and $b_{cl}=3.28$ 
($\sigma_8=0.74$); this variant of CDM is ruled
out again by direct measurements of the Hubble constant.

The standard MDM model
($\Omega_m=1$, $h=0.5$, $\Omega_b=0.5$, $n=1$, $\Omega_{\nu}=0.2$, $N_{\nu}=1$
with a best value of the cluster biasing parameter 
$b_{cl}=2.74$ ($\sigma_8=0.83$)) does significantly better: it has 
$\chi^2_{min}=23.1$ ($99\%$ C.L.) which is out of the
$2\sigma$ confidence contour but inside $3\sigma$. With the
nucleosynthesis  constraint the situation does not change:
$\chi^2_{min}=22$; also if we leave $h$ as free parameter: 
$\chi^2_{min}=21$, $h=0.48$. But if, in addition, we let vary
$\Omega_{\nu}$, we obtain $\chi^2_{min}=13$
with best-fit values of $\Omega_{\nu}=0.09$, $h=0.43$, 
$b_{cl}=3.2$ ($\sigma_8=0.73$).
This  means that the model is ruled out (as well as the model 9th in Table 4)
by the data set considered in this work at $\sim 70\%$ confidence
level only. But also here the 
best-fit value for $h$ is very low. If we fix it at lower observational
limit $h=0.5$ then $\chi^2_{min}=18.9$ (the best fit values are: 
$\Omega_{\nu}=0.15$, $b_{cl}=2.8$ ($\sigma_8=0.83$)), which
corresponds to a confidence level of 95\% .

Therefore, we conclude
that the observational data set used here rules out  CDM models with 
$h\ge 0.5$, a scale invariant primordial
power spectrum ($n=1$) and $\Omega_k=\Omega_{\Lambda}=0$
at very high confidence level, $>99.99\%$. MDM models
with $h\ge 0.5$, $n=1$ and $\Omega_k=\Omega_{\Lambda}=0$
are ruled out at $\sim 95\%$ C.L.

The best-fit parameters for 31 models which are inside of
$1\sigma$ range of the best model are presented in Table 4.
We conclude also that the observational data set used here does
not rule out any  of the 32 models presented in Table 4 at high 
confidence level but defines the  1$\sigma$ range of cosmological 
parameters for the $\Lambda$MDM models
which match observations best.


\begin{table*}
\caption{Parameters determined for the tilted $\Lambda$MDM  with
one sort of massive neutrinos if some of the data are
excluded from the searching procedure.}
\def\onerule{\noalign{\medskip\hrule\medskip}}
\medskip
\begin{tabular}{|cccccccc|}
\hline
&&&&&&&\\
Excluded data     &$\chi^2_{min}$  &$n$  & $\Omega_m$& $\Omega_{\nu}$&$\Omega_b$ & $h$   & $b_{cl}$ \\ [4pt]
\hline
&&&&&&&\\
All points of $\tilde P_{A+ACO}(k_j)$                         &1.45&1.10$\pm$0.08&0.42$\pm$0.11&0.053$\pm$0.021&0.041$\pm$0.013&0.68$\pm$0.11&****	   \\
$\tilde \ell_p$, $\tilde A_p$                                 &4.22&1.15$\pm$0.15&0.40$\pm$0.11&0.065$\pm$0.033&0.040$\pm$0.016&0.69$\pm$0.14&2.20$\pm$0.33 \\
$\tilde \sigma_8\tilde\Omega_{m}^{0.46-0.09\Omega_m}$         &3.68&1.12$\pm$0.09&0.47$\pm$0.15&0.077$\pm$0.036&0.042$\pm$0.016&0.67$\pm$0.13&2.19$\pm$0.33 \\
$\tilde \sigma_8\tilde\Omega_{m} ^{0.29}$                     &4.03&1.11$\pm$0.10&0.40$\pm$0.11&0.052$\pm$0.031&0.041$\pm$0.015&0.68$\pm$0.12&2.32$\pm$0.35 \\
Both $\sigma_8$ tests                                         &3.65&1.12$\pm$0.10&0.46$\pm$0.17&0.072$\pm$0.048&0.042$\pm$0.016&0.67$\pm$0.13&2.22$\pm$0.38 \\
$\tilde V_{50}$                                               &4.57&1.11$\pm$0.10&0.41$\pm$0.11&0.057$\pm$0.030&0.039$\pm$0.014&0.70$\pm$0.12&2.25$\pm$0.34 \\
$\tilde \sigma_{F}(z=3)$                                      &4.61&1.13$\pm$0.11&0.39$\pm$0.12&0.056$\pm$0.030&0.038$\pm$0.014&0.71$\pm$0.13&2.19$\pm$0.36 \\
$\tilde \Delta^2_{\rho}(k_p,z=2.5)$,$\tilde n_{p}(k_p,z=2.5)$ &4.41&1.13$\pm$0.10&0.41$\pm$0.11&0.069$\pm$0.035&0.038$\pm$0.014&0.70$\pm$0.13&2.19$\pm$0.36 \\
Both Ly$\alpha$ tests                                         &3.70&1.11$\pm$0.10&0.56$\pm$0.22&0.222$\pm$0.291&0.042$\pm$0.017&0.67$\pm$0.13&2.27$\pm$0.40 \\
$\tilde h$                                                    &4.28&1.11$\pm$0.10&0.35$\pm$0.13&0.051$\pm$0.024&0.030$\pm$0.017&0.79$\pm$0.22&2.11$\pm$0.39 \\
$\tilde \Omega_bh^2$                                          &4.04&1.14$\pm$0.09&0.37$\pm$0.10&0.068$\pm$0.017&0.000$\pm$0.000&0.66$\pm$0.10&2.14$\pm$0.33 \\ [4pt]
\hline
\end{tabular}
\end{table*}

One important question is how each point of the data influences our
result. To estimate this we have
excluded some data points from the searching procedure. We have determined the
best-fit parameters for the cases: 
\begin{itemize}
\item all points of Abell-ACO power
spectrum $\tilde P_{A+ACO}(k_j)$ are excluded, 
\item data on position and amplitude of
acoustic peak, $\tilde \ell_p$, $\tilde A_p$ are excluded, 
\item the value for $\sigma_8$ from \cite{gir98}, 
$\tilde \sigma_8\Omega_{m}^{0.46-0.09\Omega_m}$ is excluded, 
\item  the value for
$\sigma_8$ from  \cite{bah98}, $\tilde \sigma_8\Omega_{m} ^{0.29}$
is excluded, 
\item both these tests are excluded,
\item the bulk motion, $\tilde V_{50}$, is excluded,
\item the Ly-$\alpha$ constraint by \cite{gn98} $\tilde \sigma_{F}(z=3)$
is excluded, 
\item the Ly-$\alpha$ constraint by \cite{cr98} $\tilde
\Delta^2_{\rho}(k_p,z=2.5)$ and $\tilde n_{p}(k_p,z=2.5)$ are
excluded, 
\item both Ly-$\alpha$ tests are excluded, 
\item data on the direct measurements
of Hubble constant $\tilde h$ is excluded, and 
\item the nucleosynthesis
constraint by \cite{bur99} is not used. 
\end{itemize}
The results for models with
$N_{\nu}=1$ and all parameters free are presented in Table 7 (see for
comparison model 1 for $N_{\nu}=1$ in Table 4).
Excluding any part of observable data results only in a  change
of the best-fit values of $n$, $\Omega_m$ and $h$ within the range of
their corresponding
standard errors. This indicates that the data are mutually in
agreement, implying the same cosmological parameters (within the still
considerable error bars).
The small scale constraints, the Ly-$\alpha$ tests reduce the hot dark
matter content from
$\Omega_{\nu}\sim 0.22$ to $\sim 0.075$. The $\sigma_8$-tests further reduce
$\Omega_{\nu}$ to $\sim 0.06$. Including of the Abell-ACO power
spectrum in the search
procedure, tends to enhance  $\Omega_{\nu}$ slightly.
The most crucial test for the baryon content is of course the nucleosynthesis
constraint. Its $\sim 6\%-1\sigma$-accuracy safely keeps
$h^2\Omega_b$ near its median value 0.019. The parameter $\Om_b$ in
turn is only known to $\sim 36\%$ accuracy due to the large
errors of other experimental data used here, especially Hubble constant.
The obtained accuracy of $h$ ($\sim17\%$) is better than the one
assumed from direct measurements, $\sim 23\%$.
 Summarizing, we conclude that all data points used here
are important for searching the best-fit cosmological parameters.

\section{Discussion}

The best-fit parameters obtained in this paper are within the allowed
range of parameters found by other authors using different constraints. For
example, for the $\Lambda$MDM model with scale-invariant primordial
power spectrum and one sort of massive neutrinos which contributes 
10--20\% of matter density, \cite{vkn98} found $0.45\le \Omega_m\le
0.75$ for $h=0.5$, and $0.3\le \Omega_m\le 0.5$ for $h=0.7$. Similar
constraints have been given by \cite{pr98} for $\Lambda$MDM models
with two species of massive neutrinos.  Our values of $\Omega_m$ are
within these ranges. But at the boundary of this parameter range
$\chi^2\ge 20$, which is outside of the $2\sigma$ confidence contour.

Recently \cite{bah99} have shown that the CMB anisotropy data, the cluster
evolution and the SNIa magnitude-redshift relation indicate
a flat Universe with accelerated expansion, compatible with a
$\Om_\La \simeq 0.7$
and $\Omega_m\approx 0.3$ if CDM ($\Omega_{\nu} = 0$) is
assumed. As we can see from Table 4 (case No 11, 1), our analysis leads
to the same conclusion if we set density of hot dark
matter to the minimum value compatible with
the Super-Kamiokande experiment  (less than 1\% of  the total
density).  However, if $\Omega_\nu$ is a free
parameter, the observational data considered in this work lead to a
$\Lambda$MDM with a
slightly blue spectrum of primordial fluctuations (case No 1 in Table
4).

In our preferred tilted $\Lambda$MDM models (case No 1) the masses of
neutrinos are $m_{\nu}=2.7\pm 1.2$ eV for model with $N_{\nu}=1$,
$m_{\nu}=2.4\pm 1.0$ eV when $N_{\nu}=2$ and $m_{\nu}=2.0\pm 0.8$ eV
for model with $N_{\nu}=3$. The accuracy of neutrino mass or density
determination is modest because the observational constraints
depend stronger on
$\Omega_m$ and $n$ than on $\Omega_{\nu}$ and $N_{\nu}$. In models
with fixed low matter density $\Omega_m=0.3$ (case No 7 and 10)  the
best-fit values of the neutrino density are $\Omega_{\nu} \approx 0$,
i.e. even below the lower limit of the massive neutrino contribution to
the cosmological density indicated by the Super-Kamiokande
experiment. However, the $1\sigma$ contours of the low $\Omega$ models
include the Super-Kamiokande limit (see Fig.~\ref{Lcm7}b).

In the last column of Table 5 we also indicate the age of the Universe,
\be
t_0={2\over 3H_0}\left[{1\over
2\Omega_{\Lambda}^{1/2}}\ln{1+\Omega_{\Lambda}^{1/2}\over
1-\Omega_{\Lambda}^{1/2}}\right],
\ee
for each model as well as the age of the oldest globular
clusters (\cite{car99}).  All models with parameters taken from
Table 4 have ages which are in  agreement with the oldest objects of our
galaxy.

We have used a scale independent, linear bias factor $b_{cl}$ as free
parameter in order to fit the Abell-ACO power spectrum amplitude.

Let us discuss in more detail how the model predictions presented in
Table 5 match each observable constraint separately. The predicted
position of the acoustic peak for all models is lower than the one
determined from the observational data set
presented in Table 2  ($\ell_p=253\pm 70$).
Tilted $\Lambda$MDM models prefer $\ell_p\sim 210-230$.
This is due to the fact that the peak position depends only very
weakly on the parameters discussed in this work. It is determined
mainly by spatial curvature which we have set to zero here (together
with the initial conditions which we have assumed to be adiabatic). 
However, our result is in good agreement with the most recent and so
far most accurate determination of the peak position from one single
experiment, the  North American test flight of
Boomerang (\cite{mau99,mel99b}),  which led to
$0.85\le\Om_m + \Om_\La\le 1.25$ with maximum likelihood near 1 for
adiabatic CDM models. The prediction of our best model for position
of the first acoustic peak ($\ell=215$)  matches the value given
by Boomerang experiment $\ell\sim 200$ very well.
The central value from the combination of all available experiments,
$\ell_p=253$, may very well be contaminated by mutual calibration 
inconsistencies. 

Finally we want to discuss the possibility of using the averaged power
spectrum of galaxies obtained by \cite{ein99} to determine the
parameters. 
 This averaged spectrum of galaxies is determined in a wide
range of scales ($0.02h/$Mpc$\le k\le 10h/$Mpc) and has substantially lower
errors than the Abell-ACO power spectrum used here.  Its
$1\sigma$ errors are $\sim 4\%$  on small scales
and $\sim 20\%$ at
large scales versus $\sim 40\%$ and $\sim 60\%$ respectively for the
Abell-ACO power spectrum. It is interesting to compare the predictions
obtained from the power spectrum of galaxies with our analysis,
because, as already mentioned in the introduction, the correction of
the linear power spectrum for nonlinear
evolution must be included into the algorithm. We use the
fitting function by \cite{sm97}, which transfers the linear 
into the nonlinear power spectrum, and the observational constraint for
the Hubble constant $\tilde h=0.6\pm 0.02$ (\cite{Saha99},\cite{Tamm99})
as well as the nucleosynthesis constraint for the baryon content by
\cite{bur99}. Under these assumptions we find  the following
 best fit parameters: $n=1.11\pm 0.02$, $h=0.62\pm 0.02$,
$\Omega_m=0.2\pm 0.03$, $\Omega_b=0.044\pm 0.005$,
$\Omega_{\nu}=0.01\pm 0.01$, $N_{\nu}=3$ and galaxy biasing
parameter $b_g=1.52\pm 0.06$.

If we add the remaining observations described in Sect.s 2.2 and 2.3
the best-fit parameters remain practically unchanged due to the large
number of (probably not independent) data points in the galaxy power
spectrum. A model with these parameters has serious
problems reproducing the  experimental data set used here. Indeed, with
these parameters we obtain  $\chi^2\approx 61$, for the data set used
in the rest of this work, far outside 3$\sigma$ contour.
The model predictions
$\sigma_8\Omega_{m}^{0.46-0.09\Omega_m}=0.28$ and $\sigma_8\Omega_{m}
^{0.29}=0.35$ are $\sim 4\sigma$ lower than the corresponding
observational values by \cite{gir98} and \cite{bah98}.
Moreover, the peculiar velocity $V_{50}$ is $\sim 2\sigma$
lower than the observed value, $\sigma_{F}(z=3)$ is  and
$\Delta^2_{\rho}(k_p,z=2.5)$ are $\sim 3\sigma$  and $\sim 1.5\sigma$ 
lower than the corresponding values inferred from the Ly-$\alpha$ 
measurements. Therefore, we conclude that a model with parameters
determined by the galaxy power spectrum is ruled out by the
observations discussed in this work.

This result is not completely unexpected, because the galaxy power
spectrum on small scales is probably influenced by a scale dependent
bias (see for example \cite{kra99}, Fig. 3) which is not taken into
account here. Moreover, the fitting formula for nonlinear
evolution at $k\ge$1 h/Mpc may be incorrect. If we disregard the short
wavelength part of galaxy power spectrum we find
parameters close to those presented in Table 4.

\section{Conclusions}

Using Levenberg-Marquardt $\chi^2$ minimization method we have
determined the cosmological parameters of spatially flat, tilted $\Lambda$MDM models.
We searched for a maximum of 6 parameters: the spectral index $n$,
the matter content $\Omega_m$ ($\Omega_m+\Omega_{\Lambda}=1$), the hot dark
matter content $\Omega_{\nu}$, the baryon content $\Omega_b$,
the dimensionless Hubble constant $h$ and the  biasing parameter for
rich clusters, $b_{cl}$. The experimental data set used in the search procedure
included the Abell-ACO power spectrum (\cite{ret97}), the position and
amplitude of the first acoustic peak in the angular power spectrum of CMB
temperature fluctuations determined from the set of published
measurements on different scales, the constraints for the density
fluctuation amplitude $\sigma_8$ derived from the mass function of nearby and
distant clusters (\cite{gir98,bah98}), the mean peculiar velocity of
galaxies in a sphere of radius $50h^{-1}$Mpc (\cite{kol97}), the
constraints on amplitude and tilt of the power spectra at small
scales obtained from Ly-$\alpha$ clouds at z=2-3 (\cite{gn98,cr98}), the
nucleosynthesis constraints (\cite{tyt96,bur99}) and the COBE data
(\cite{bun97}) which is used to  normalize the  model power spectra.

We have considered one, two and three species of
massive neutrinos. We have studied the influence of a reduction of the
number of free parameters.  In Table 4 we summarize the parameters
which we have determined in 33 different cases. Based on the results
presented in Table 4 we conclude:

\begin{itemize}

\item The tilted $\Lambda$MDM model with one sort of massive
neutrinos and the best-fit parameters $n=1.12\pm 0.10$,
$\Omega_m=0.41\pm 0.11$, $\Omega_{\Lambda}=0.59\pm0.11$, $\Omega_{\nu}=0.059\pm 0.028$,
$\Omega_b=0.039\pm 0.014$ and $h=0.70\pm 0.12$ (standard errors) matches the
observational data set best.
The 1$\sigma$ (68.3\%) confidence limits on each cosmological
parameter, obtained by marginalizing over the other parameters, are
 $0.82\le n\le 1.39$, $0.19\le\Omega_m\le 1$, $0\le\Omega_{\Lambda}\le0.81$,
$0\le\Omega_{\nu}\le 0.17$, $0.021\le\Omega_b\le 0.13$ and
$0.38\le h\le 0.85$.

\item The degeneracies in the $n-\Omega_m$ and $\Omega_{\nu}-\Omega_m$ planes
 $n\sqrt{\Omega_m}=0.73$ and $\Omega_{\nu}=0.023-0.44\Omega_m+1.3\Omega_m^2$
  are revealed.

\item For fixed Hubble constant $h$ raising from 0.5 to 0.72,
the best-fit value for $\Omega_m$ decreases from 0.63  to 0.39 for 
$\Lambda$MDM models with $N_{\nu}=1$. For models
with $N_{\nu}=2$ and 3 the value of $\Omega_m$ raises by $\sim 0.08$ and
$\sim 0.15$ respectively. The  $\Omega_{\nu}$ is higher for more species
of massive neutrinos, $\sim0.06$ for one sort and $\sim0.13$ for three, and
decreases slowly for growing $h$. The inclination of initial power spectrum
$n$ correlates somewhat with $\Omega_{\nu}$ and grows slightly with $h$.

\item Fixing a low $\Omega_m$=0.3 a $\Lambda$CDM model without HDM
 matches the observational data set best. In this case the parameters
are $h=0.71\pm 0.12$, $n=1.04\pm 0.10$ and $\Omega_b=0.038\pm 0.013$.

\item For all models the biasing parameter $b_{cl}$ of rich clusters
is in the range 2.2-3.3, for the best model it equals $2.23\pm 0.33$
(standard error). The 1$\sigma$ (68.3\%) confidence interval is
$1.5\le b_{cl}\le 3.5$.

\item  CDM models with $h\ge 0.5$, scale invariant
primordial power spectrum $n=1$ and $\Omega_{\Lambda}=\Omega_k=0$
are ruled out at very high confidence level, $>99.99\%$.

\item Also pure MDM models are ruled out at $\sim 95\%$ C.L.

\end{itemize}

Finally, we note that the accuracy of present
observational data on the large scale structure of the Universe is
still too low to constrain the  set of cosmological parameters
sufficiently, but we believe that our work shows the potential of such
studies, which search for parameters including data from different,
often complementary observations. It is clear that with sufficiently
accurate data, such a study may also reveal an inconsistency of model
assumptions.

{\it Acknowledgments} This work is part of a project supported by the
Swiss National Science Foundation (grant NSF 7IP050163).
B.N. is also grateful to DAAD for financial support (Ref. 325)
and AIP for hospitality. V.N.L. is grateful to  INTAS support
(97-1192).

\end{document}